\documentclass{cernyrep}
\usepackage{texnames}
\usepackage{fancyhdr}
\usepackage[colorlinks=true, linkcolor=black, citecolor=black, filecolor=black, urlcolor=blue]{hyperref}
\fancyhfoffset{4 mm}
\fancypagestyle{ARTTITLE}{%
\fancyhf{} 
\lhead{\small{Proceedings of the CERN--Accelerator--School course: \it{Introduction to Accelerator Physics}}}
\lfoot{Available online at \url{https://cas.web.cern.ch/previous-schools}}
\rfoot{\thepage\hspace*{3mm}}

}

\usepackage[T1]{fontenc}
\usepackage{comment}
\begin{document}
\title{A first taste of nonlinear beam dynamics}
 
\author{Hannes Bartosik$^\dagger$, Yannis Papaphilippou$^\dagger$, Andrzej Wolski$^\ddag$}

\institute{$^\dagger$CERN, Geneva, Switzerland,\\$^\ddag$University of Liverpool, and the Cockcroft Institute, UK}

\maketitle 
\thispagestyle{ARTTITLE}

\begin{abstract}
Nonlinear dynamics can impact the performance of a particle accelerator in a number
of different ways, depending on the type of the accelerator and the parameter regime in
which it operates.  Effects can range from minor changes in beam properties or
behaviour, to serious limitations on beam stability and machine performance.  In these
notes, we provide a brief introduction to nonlinear dynamics in accelerators.  After a review of
some relevant results from linear dynamics, we outline some of the main ideas of
nonlinear dynamics, framing the discussion in the context of two examples of different
types of accelerator: a single-pass system (a bunch compressor) and a periodic system
(a storage ring).  We show how an understanding of the origins and nature of the
nonlinear behaviour, together with the use of appropriate analysis tools, can prove
useful in predicting the effects of nonlinear dynamics in different systems, and allow
the design of appropriate corrections or mitigations.
\end{abstract}
 
\section{Introduction}

In these notes on nonlinear dynamics in particle accelerators, we shall discuss a number of topics:
\begin{itemize}
\item effects of nonlinear perturbations in single-pass and periodic systems,
including phase space distortions, resonances, tune shifts and dynamic aperture limitations;
\item mathematical tools for modelling nonlinear dynamics, including
power series maps and symplectic maps;
\item analysis methods such as normal form analysis and frequency map analysis.
\end{itemize}
Our goal is not to provide a rigorous or comprehensive review of nonlinear
dynamics in accelerators (which is a very large subject), but to provide a
short introduction to some of the key concepts and phenomena.
The topics that we discuss are covered in many publications: some suggestions
for further reading, where specific topics that may be of interest are covered in
greater depth, are given in section~\ref{summary}.
We shall frame our discussion in the context of two types of accelerator system:
first, a bunch compressor (a single-pass system), and later, a storage ring (a
multi-turn system).  To begin with, however, we briefly review some of the
principles of linear dynamics in particle accelerators, that form an important
foundation for the development of ideas in nonlinear dynamics.


\section{Foundations from linear dynamics}

Particle motion through simple components such as drifts, dipoles and quadrupoles
can be represented by \emph{linear transfer maps}\footnote{Linear dynamics is covered
in many standard texts in accelerator physics, for example \cite{wiedemann2015,sylee2011,wolski2014}.}.
For example, in a drift space, the horizontal co-ordinate $x$ and momentum $p_x$
change from initial values $x_0$ and $p_{x0}$ at the entrance to the drift space, to
final values $x_1$ and $p_{x1}$ at the exit, given by:
\begin{eqnarray}
x_1 & = & x_0 + L p_{x0}, \label{driftmap1} \\
p_{x1} & = & p_{x0}, \label{driftmap2}
\end{eqnarray}
where $L$ is the length of the drift space.
Note that the horizontal (canonical) momentum is given by:
\begin{equation}
p_x = \frac{\gamma m v_x}{P_0},
\end{equation}
where $\gamma$ is the relativistic factor, $m$ is the rest mass of the particle,
$v_x$ is the horizontal velocity, and $P_0$ is the reference momentum.
The reference momentum is a fixed value (in the absence of acceleration)
that is used mainly for convenience, for scaling quantities such as the particle
momentum.  In principle, the reference momentum can be chosen arbitrarily,
though it is usually best to choose a value equal to the ``design'' momentum
of the beam.
For small values of $p_x$ (i.e.~for $|p_x \ll 1|$), the horizontal momentum is
approximately equal to the angle of the trajectory with respect to the design
trajectory:
\begin{equation}
p_x \approx \frac{dx}{ds}.
\end{equation}
The reference trajectory is simply a line through space that acts as the origin
of the local cartesian co-ordinate system, with (transverse) axes $x$ horizontally
and $y$ vertically.  The variable $s$ is the distance along the reference trajectory
from a fixed starting point.

Linear transfer maps can be written in terms of matrices.
For example, for a drift space of length $L$, the map given by equations
(\ref{driftmap1}) and (\ref{driftmap2}) can be written:
\begin{equation}
\left( \begin{array}{c}
x_1 \\ p_{x1}
\end{array} \right)
= \left( \begin{array}{cc}
1 & L \\
0 & 1
\end{array} \right)
\left( \begin{array}{c}
x_0 \\ p_{x0}
\end{array} \right).
\end{equation}
In general, a linear transformation can be written:
\begin{equation}
\vec{x}_1 = R \, \vec{x}_0 + \vec{A}, \label{linearmapgeneral}
\end{equation}
where $\vec{x}_0$ and $\vec{x}_1$ are the initial and final
phase space vectors, with components $(x_0, p_{x0})$ and
$(x_1, p_{x1})$, respectively,
$R$ is a matrix (the \emph{transfer matrix}) and $\vec{A}$ is a vector.
The components of $R$ and $\vec{A}$ are constant, i.e.~they do not depend
on $\vec{x}_0$: it is this feature that makes the map (\ref{linearmapgeneral})
a \emph{linear} map.

In the case that $\vec{A} = 0$ for all elements in a beamline, the transfer matrix
for that beamline can be found simply by multiplying
the transfer matrices for the accelerator components within that beamline.  For
example, if a beamline consists of a sequence of elements with transfer matrices
$R_1$, $R_2$, $R_3$ $\ldots$ $R_n$, with the beam passing through the elements
in that order, then after the first element, the phase space vector becomes:
\begin{equation}
\vec{x}_1 = R_1 \vec{x}_0.
\end{equation}
After the second element, it becomes:
\begin{equation}
\vec{x}_2 = R_2 \vec{x}_1 = R_2 R_1 \vec{x}_0.
\end{equation}
Eventually, after $n$ elements, the phase space vector is:
\begin{equation}
\vec{x}_n = R_\mathrm{total} \vec{x}_0 = R_n R_{n-1} \cdots R_2 R_1 \vec{x}_0.
\end{equation}
The transfer matrix $R_\mathrm{total}$ is constructed by multiplying the transfer
matrices for the individual elements, writing them in the multiplication in the reverse
order that they appear in the beamline.

For a periodic beamline (i.e.~a beamline constructed from a repeated unit)
the transfer matrix for a single period can be parameterised in terms of the Courant--Snyder
parameters $(\alpha_x, \beta_x, \gamma_x)$ and the phase advance, $\mu_x$:
\begin{equation}
R = \left( \begin{array}{cc}
\cos(\mu_x)+\alpha_x \sin(\mu_x) & \beta_x \sin(\mu_x) \\
-\gamma_x \sin(\mu_x) & \cos(\mu_x) - \alpha_x \sin(\mu_x)
\end{array} \right). \label{transfermatrixperiodicbeamline}
\end{equation}
If the transfer matrix is found by multiplying together the transfer matrices for
the elements in a unit cell (in the appropriate order), then the values of
the Courant--Snyder parameters and the phase advance $\mu_x$ can be
found by equating the components of the matrix in terms of the beamline elements
with the components of the matrix given in the form (\ref{transfermatrixperiodicbeamline}).

Neglecting synchrotron radiation and various collective effects, the transfer matrices
will be \emph{symplectic}.  For a $2\times 2$ matrix, this means that the matrix will
have unit determinant, which leads to the condition:
\begin{equation}
\beta_x \gamma_x - \alpha_x^2 = 1.
\end{equation}

\begin{figure}
\begin{center}
\includegraphics[width=0.8\textwidth]{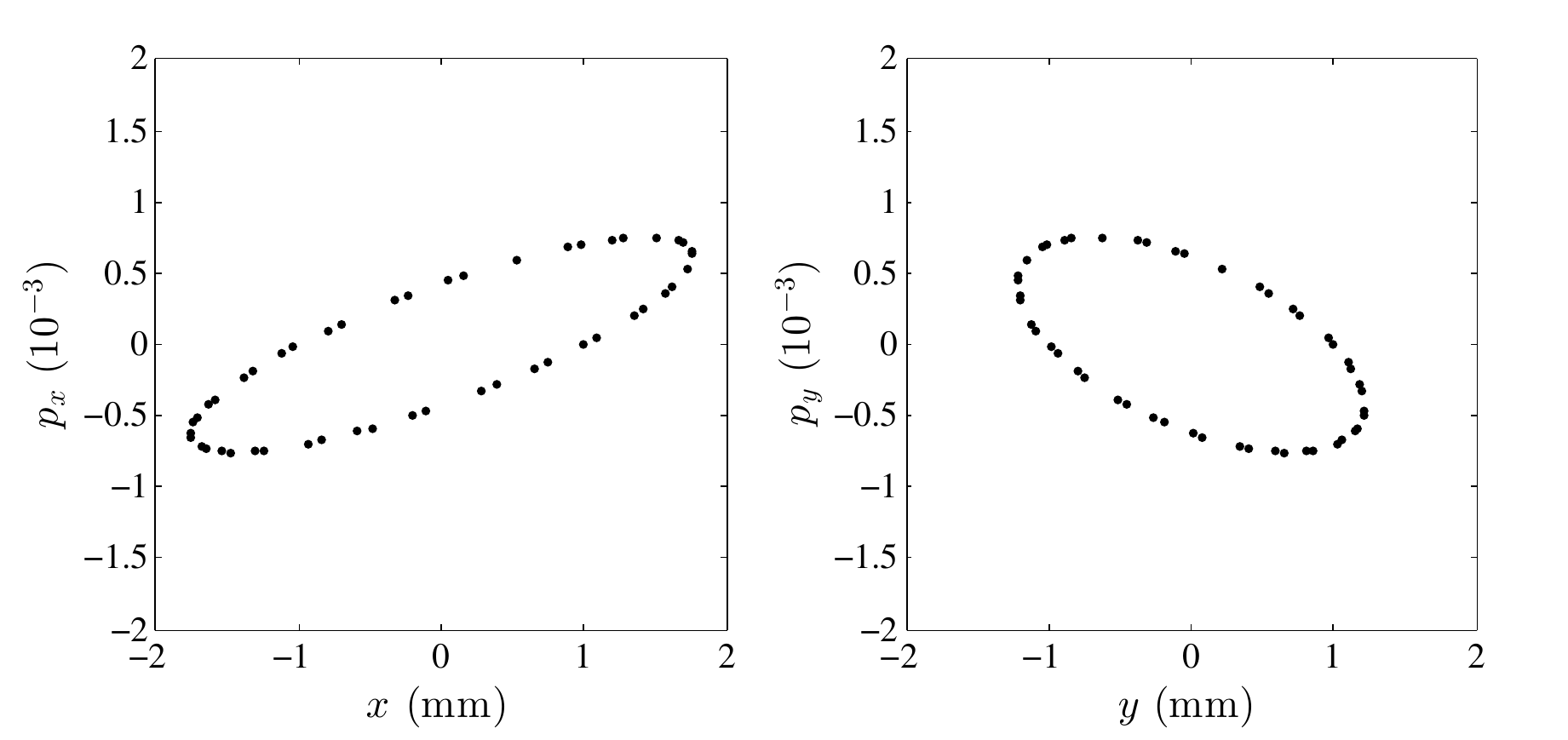}
\end{center}
\caption{Horizontal and vertical phase space portraits for a particle
moving along a periodic beamline.  Each point indicates the co-ordinate
and corresponding component of the momentum of the particle after
each full pass through a periodic section of the beamline.  Since the
beamline is linear and stable, the points lie on ellipses with shapes
determined by the Courant--Snyder parameters for the beamline.  The
area of the ellipse is determined by the betatron amplitude of the particle
in each plane.  The angle around the ellipse that the particle moves in
each period corresponds to the phase advance.
\label{phasespaceellipses}}
\end{figure}

\begin{figure}
\begin{center}
\includegraphics[width=0.5\textwidth]{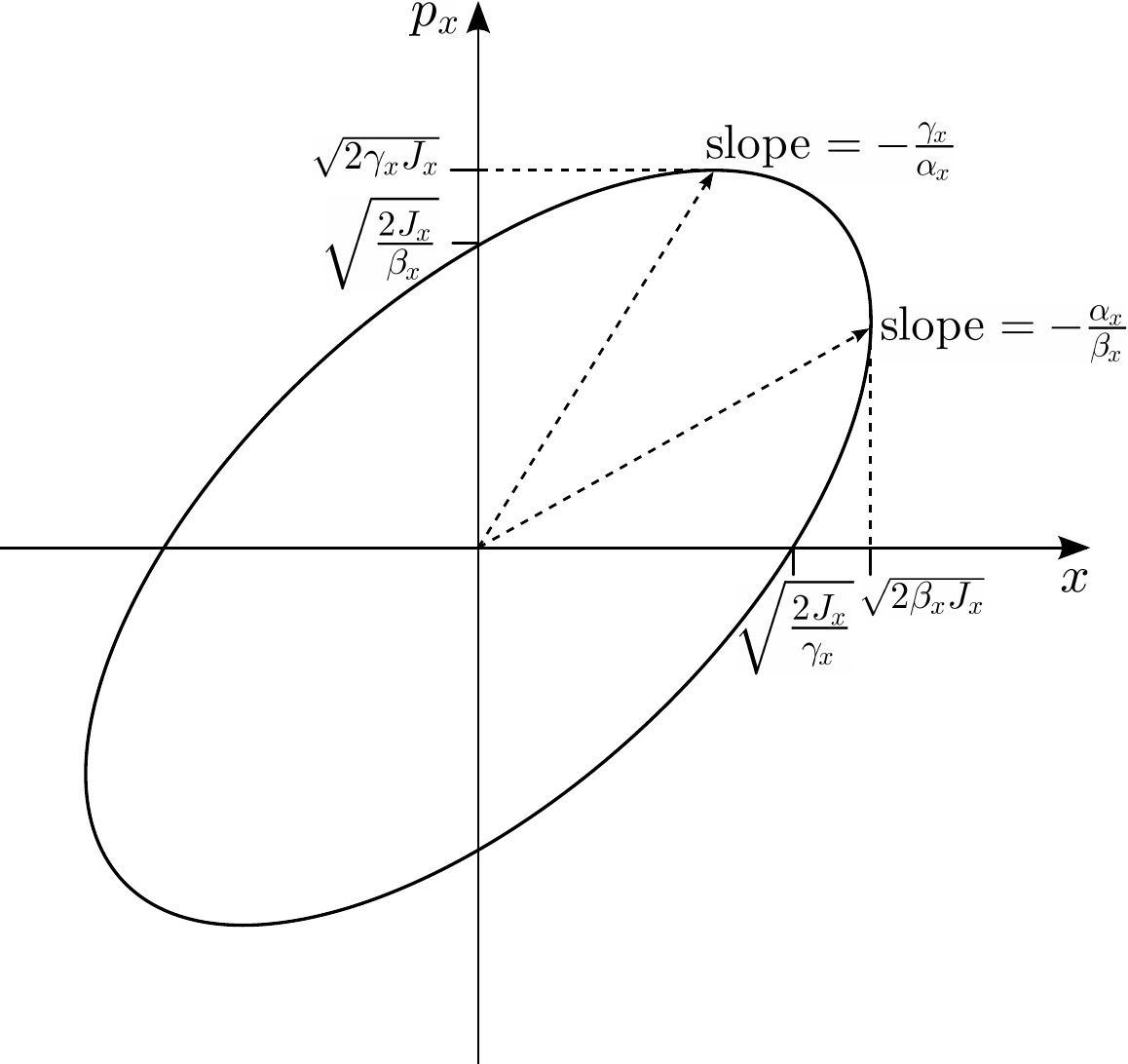}
\end{center}
\caption{The Courant--Snyder parameters describe the shape of the ellipse
mapped out in phase space as a particle moves through successive periods
in a periodic beamline.  The area of the ellipse, corresponding to the betatron
amplitude, is $\pi J_x$, where
$J_x$ is the betatron action.
\label{courantsnyderellipse}}
\end{figure}

If the beamline is stable, then the characteristics of the particle motion can
be represented by a \emph{phase space portrait} showing the co-ordinates and
momenta of a particle after an increasing number of passes through successive
periods of the beamline.
If the transfer map for each period is linear, then the phase space
portrait is an ellipse with area $\pi J_x$: examples (for horizontal and vertical
motion) are shown in Fig.~\ref{phasespaceellipses}.  The quantity $J_x$ is called
the \emph{betatron action}, and characterises the amplitude of the betatron oscillations.
The shape of the ellipse is described by the Courant--Snyder parameters, as shown in
Fig.~\ref{courantsnyderellipse}. The rate at which particles move around the ellipse
corresponds to the phase advance in each periodic secton of the beamline, and (for
linear motion) is independent of the betatron action.

The betatron action $J_x$ and the phase (or angle) $\phi_x$ provide an alternative to the
regular cartesian variables $x$, $p_x$ for describing the motion of particles in an
accelerator beamline.
The cartesian variables can be expressed in terms of the action--angle variables:
\begin{eqnarray}
x & = & \sqrt{2\beta_x J_x} \cos(\phi_x), \label{xfromjphi} \\
p_x & = & - \sqrt{\frac{2J_x}{\beta_x}}(\sin(\phi_x) + \alpha_x \cos(\phi_x)). \label{pxfromjphi}
\end{eqnarray}

\section{From linear to nonlinear maps}

Nonlinearities in the particle dynamics can come from a number of different sources,
including:
stray fields in drift spaces;
higher-order multipole components in dipoles and quadrupoles;
higher-order multipole magnets (sextupoles, octupoles etc.) used
to control various properties of the beam;
effects of fields generated by a bunch of particles on individual
particles within the bunch (space-charge forces, beam-beam effects, and many others).
The extent to which each of these (or other) sources contributes to nonlinear effects
in an accelerator depends very much on the specific case.

The effects of nonlinearities can be varied and quite dramatic, and can impact the
operation and overall performance of an accelerator in may different ways.  It is important
to have some understanding of nonlinear dynamics for optimising the design
and operation of many accelerator systems.  It is also important to have the appropriate
mathematical tools and techniques for the analysis of nonlinear dynamics, and a range
of powerful methods have been developed over the years.  We shall mention a few of these
methods in a later section, but one of simplest approaches is to write the transfer map as a power
series in the dynamical variables: this is a natural extension of linear transfer maps,
given (for example) in a drift space in Eqs.~(\ref{driftmap1}) and (\ref{driftmap2}).
As an example of this approach, consider a particle moving through a sextupole magnet.
The vertical component of the field in a sextupole is given by:
\begin{equation}
\frac{B_y}{B\rho} = \frac{1}{2} k_2 x^2, 
\end{equation}
where $B\rho = P_0/q$ is the beam rigidity, and $k_2$ is the normalised sextupole gradient.
In the ``thin lens'' approximation, the horizontal deflection of a particle on passing through the
sextupole is:
\begin{equation}
\Delta p_x = - \frac{1}{B\rho} \int B_y \, ds \approx - \frac{1}{2}k_2 L x^2,
\end{equation}
where $L$ is the length of the sextupole.
Hence, the transfer map for a sextupole in the thin lens approximation is:
\begin{eqnarray}
x_1 & = & x_0, \\
p_{x1} & = & p_{x0} - \frac{1}{2}k_2 L x_0^2.
\end{eqnarray}
Although we can write the effect of the sextupole as a transfer map in the same
way that we did for a drift space in Eqs.~(\ref{driftmap1}) and (\ref{driftmap2}),
in the case of the sextupole, the map is \emph{nonlinear}: it depends on higher-order
terms in the original values of the phase space variables.  This means that the
transfer map cannot be written simply as a matrix.  However, we can generalise
the matrix equation (\ref{linearmapgeneral}), to express a nonlinear transfer map
as a power series:
\begin{eqnarray}
x_1 & = & A_1 + R_{11} x_0 + R_{12} p_{x0} + T_{111}x_0^2 + T_{112} x_0 p_{x0} + T_{122} p_{x0}^2 + \ldots \\
p_{x1} & = & A_2 + R_{21} x_0 + R_{22} p_{x0} + T_{211}x_0^2 + T_{212} x_0 p_{x0} + T_{222} p_{x0}^2 + \ldots
\end{eqnarray}

The coefficients $R_{ij}$ correspond to components of the transfer matrix $R$.
The coefficients of higher-order (nonlinear) terms are conventionally represented
by $T_{ijk}$ (second order), $U_{ijk\ell}$ (third order) and so on.
The values of the indices correspond to the components of the phase space vector,
thus:
\begin{center}
\begin{tabular}{r|cccccc}
index value & 1 & 2 & 3 & 4 & 5 & 6 \\
\hline
component & $x$ & $p_x$ & $y$ & $p_y$ & $z$ & $\delta$
\end{tabular}
\end{center}
Hence, $T_{212}$ is the coefficient for a second-order term (three indices),
referring specifically to the part of the map for $p_x$ (first index, $i = 2$),
depending on the product of $x$ (second index, $j = 1$) and $p_x$
(third index, $k = 2$).  Because multiplication is commutative, there is no
distinction between (for example) $T_{212}$ and $T_{221}$.

The effects of nonlinearities depend very much on the type of accelerator system
being considered.  It is often useful, in this context, to make a distinction between
periodic beamlines (e.g.~in a storage ring) and non-periodic, or single-pass systems.
In a periodic beamline, the effects of nonlinearities can include:
\begin{itemize}
\item distortion of the shape of the phase space ellipse;
\item dependence of the phase advance per period on the betatron amplitude;
\item instability of motion at large amplitudes;
\item the appearance of features such as ``phase space islands'' (closed loops around points
away from the origin) in the phase space portrait.
\end{itemize}
Before considering periodic beamlines, however, we shall consider in more
detail the effects of nonlinearities in a single-pass beamline.  The discussion
(in the following section) will be based on the example of a bunch compressor.


\section{Nonlinear dynamics in a single-pass beamline: a bunch compressor\label{bunchcompressor}}

As an example of how nonlinear effects can impact the performance of an
accelerator, we shall consider a bunch compressor.
Bunch compressors reduce the length of a bunch by performing a
rotation in longitudinal phase space, and are used, for example,
in free electron lasers to increase the peak current.

We first outline the structure of the bunch compressor, and construct the
complete transfer map by combining the transfer maps for the different sections.
We then specify the parameters of the principal components based on consideration
of the linear dynamics and the compression ratio that the bunch compressor should
achieve.  This provides the basis for an analysis of both
linear and nonlinear effects: based on the results of this analysis, the final
step is to adjust the parameters, to compensate (as far as possible) the
nonlinear effects that may adversely affect the performance of the system.

\subsection{Construction of the transfer map}

\begin{figure}
\begin{center}
\includegraphics[width=0.8\textwidth]{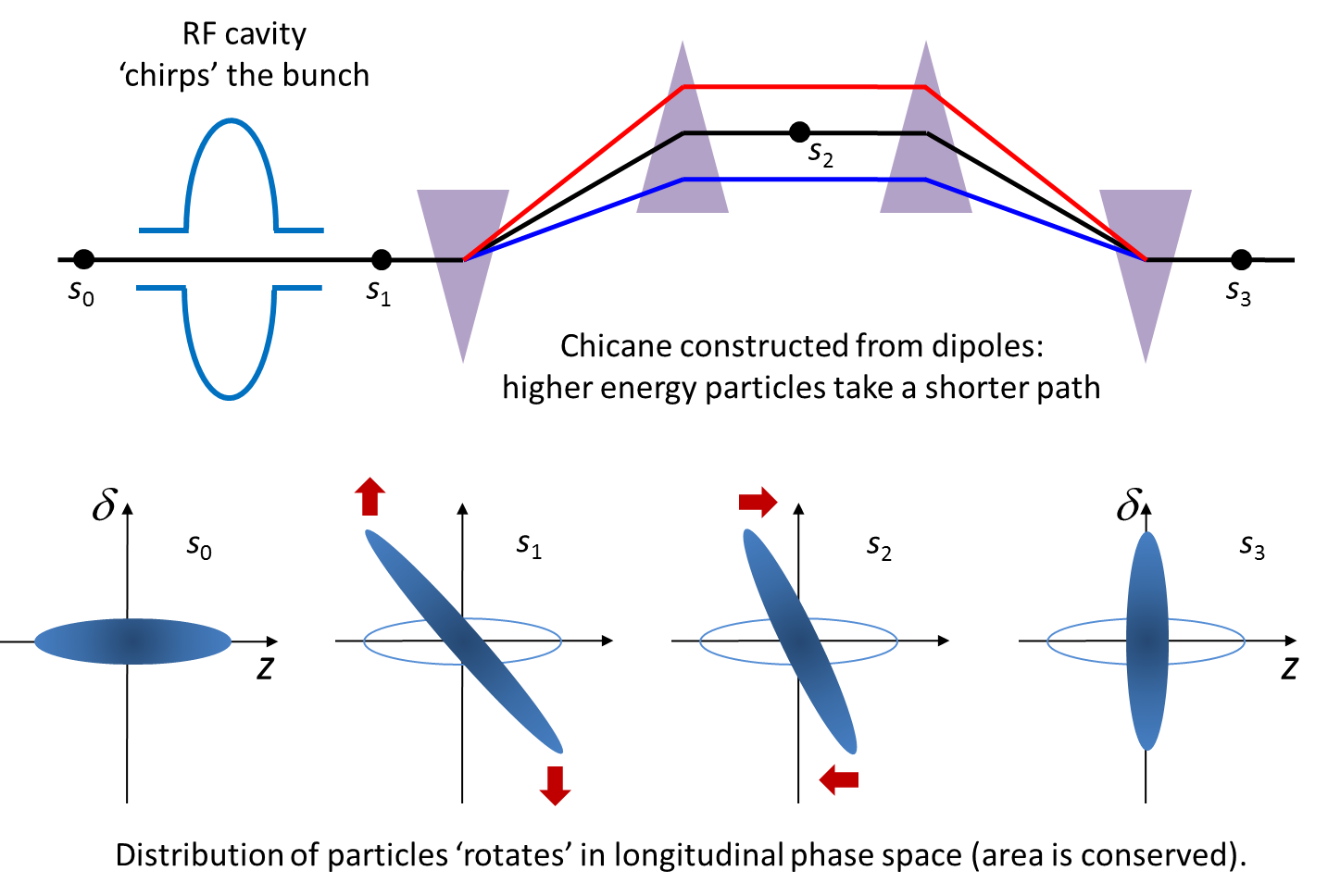}
\end{center}
\caption{Top: schematic of a bunch compressor.  A bunch of particles (moving
from left to right along the beamline) first passes through an RF cavity, and then
through a sequence of dipoles.  Bottom: changes in the distribution of particles in
the bunch in longitudinal phase space, indicating the transformations that take place
as the bunch moves through the bunch compressor.
\label{bunchcompressorschematic}
}
\end{figure}

A schematic of the bunch compressor that we shall use in this example is shown
in Fig.~\ref{bunchcompressorschematic} (top), and consists of an RF cavity
followed by a chicane constructed from four dipoles.
The RF cavity is designed to ``chirp'' the bunch, i.e.~to provide a change in
energy deviation as a function of longitudinal position $z$ within the bunch
($z>0$ at the head of the bunch).  The accelerating field in the cavity
varies with time.  Suppose that a particle in the centre of the bunch arrives
at the cavity at a time such that the average accelerating field that it sees
in the cavity is zero: the energy of this particle will be unchanged.  However,
particles ahead or behind this particle will see a non-zero field (on average)
and will then either gain or lose energy.  The energy change $\Delta E$ of
a particular particle will depend on the distance by which it is either ahead or
behind the centre of the bunch, the RF voltage and frequency of the cavity:
\begin{equation}
\Delta E = - eV \sin(k z).
\label{cavitychirp1}
\end{equation}
Here, $V$ is the maximum voltage across the cavity (taking into account the
time of flight of a particle through the cavity), $k = \omega/c$ (where $\omega$
is the RF angular frequency and $c$ is the speed of light), and $z$ is the distance of
the particle from the centre of the bunch.  Note that we use the convention that
$z$ is \emph{positive} for a particle that is \emph{ahead} of the centre of the bunch.
In that case, the minus sign in Eq.~(\ref{cavitychirp1}) means that for a bunch
of particles moving through the bunch compressor, the RF cavity reduces the
energy of particles at the head of the bunch, and increases the energy of particles
in the tail (see Fig.~\ref{bunchcompressorschematic}, bottom). 

For convenience, we define the longitudinal phase space variables as $z$ (the
co-ordinate) and $\delta$, the energy deviation.  The energy deviation $\delta$
of a particle with energy $E$ is defined as:
\begin{equation}
\delta = \frac{E - E_0}{E_0},
\label{energydeviationdefinition}
\end{equation}
where $E_0$ is the reference energy for the system (corresponding to the energy
of a particle with total momentum $P_0$, the reference momentum).
If the reference momentum is chosen so that it is close to the ``average'' momentum
of particles in the beam, then $\delta$ should be a small number (i.e.~$|\delta | \ll 1$)
for all particles: this has some advantages when expressing nonlinear transfer maps
in terms of power series of the dynamical variables.
In terms of the variables $z$ and $\delta$, the transfer map for the RF cavity in the
bunch compressor is:
\begin{eqnarray}
z_1 & = & z_0, \label{rfmap1} \\
\delta_1 & = & \delta_0 - \frac{eV}{E_0} \sin(kz_0). \label{rfmap2}
\end{eqnarray}

\begin{figure}
\begin{center}
\includegraphics[width=0.7\textwidth]{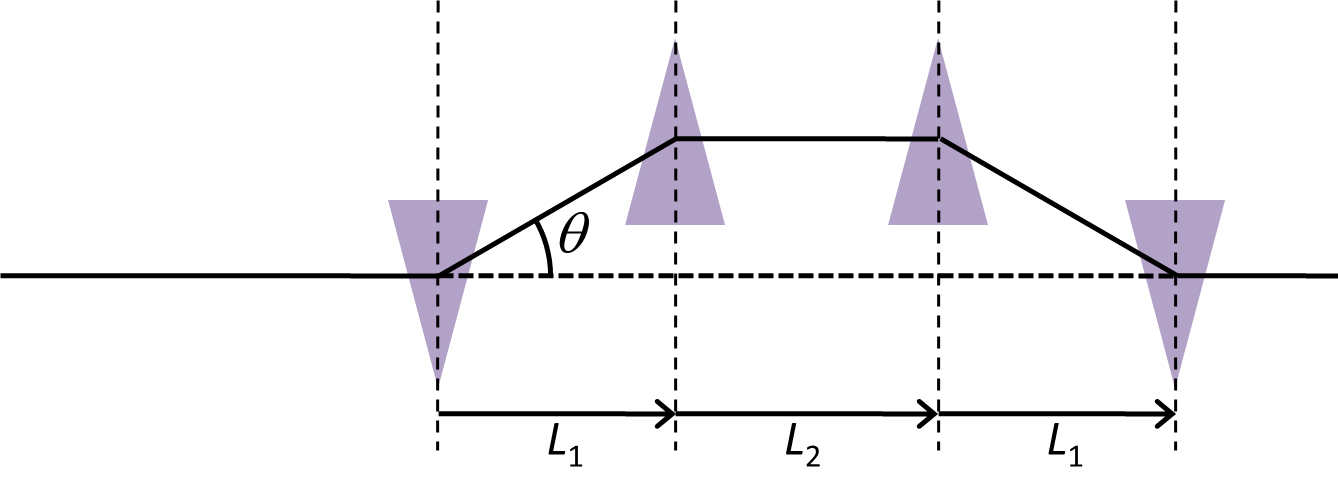}
\end{center}
\caption{Path of a particle moving through a chicane.  Since the bending
angle $\theta$ depends on the energy of a particle, the path length also
depends on $\theta$: the higher the momentum of a particle, the smaller
the angle $\theta$, and the shorter the path of the particle through the
chicane.
\label{chicanepathlength}
}
\end{figure}

Now consider the dynamics in the sequence of four dipoles following the RF cavity:
in the example we are considering here, the dipoles form a chicane, so that for a
particle with the reference momentum, the trajectory after the exit of the chicane
is colinear with the trajectory at the entrance of the chicane.  Neglecting synchrotron
radiation, the chicane does not change the energy of the particles.  However, the
path length $L$ depends on the energy of the particle.
If we assume that the bending angle in each dipole is small, $\theta \ll 1$, then from
Fig.~\ref{chicanepathlength} we find that:
\begin{equation}
L = \frac{2L_1}{\cos (\theta)} + L_2. 
\end{equation}
The bending angle is a function of the energy of the particle:
\begin{equation}
\theta(\delta) = \frac{\theta_0}{1 + \delta}. \label{theta} 
\end{equation}

Assuming that the beam is ultrarelativistic (so that we can neglect differences in 
the speed of particles arising from differences in energy) the change in the co-ordinate
$z$ of a particle after passing through the chicane is the difference between the nominal
path length, and the length of the path actually taken by the particle.
Hence, the transfer map for the chicane can be written:
\begin{eqnarray}
z_2 & = & z_1 + 2L_1 \left( \frac{1}{\cos(\theta_0)} - \frac{1}{\cos(\theta(\delta_1))} \right),
\label{chicanetransfermapdz} \\
\delta_2 & = & \delta_1,
\label{chicanetransfermapddelta}
\end{eqnarray}
where $\theta_0$ is the nominal bending angle of each dipole in the chicane, and
$\theta(\delta)$ is given by (\ref{theta}).

We can now write down the complete transfer map for the bunch compressor by
substituting $z_1$ and $\delta_1$ from Eqs.~(\ref{rfmap1}) and (\ref{rfmap2})
into Eqs.~(\ref{chicanetransfermapdz}) and (\ref{chicanetransfermapddelta}).
Expanding the resulting expressions in terms of the dynamical variables provides
the transfer map in the form of a power series.  Using the standard notation,
we write:
\begin{eqnarray}
z_2 & = & R_{55} z_0 + R_{56} \delta_0 + T_{555} z_0^2 + T_{556} z_0 \delta_0 + T_{566} \delta_0^2 + \ldots
\label{bcfullmapz} \\
\delta_2 & = & R_{65} z_0 + R_{66} \delta_0 + T_{655} z_0^2 + T_{656} z_0 \delta_0 + T_{666} \delta_0^2 + \ldots
\label{bcfullmapdelta}
\end{eqnarray}
The procedure that we have described provides expressions for the coefficients in
terms of the bunch compressor parameters.  In particular, we find for the coefficients
of the linear terms:
\begin{eqnarray}
R_{55} & = & 1 - \frac{eV}{E_0} k L_1 \frac{\theta_0 \sin(2\theta_0)}{\cos^3(\theta_0)}, \label{r55expression} \\
R_{56} & = & L_1 \frac{\theta_0 \sin(2\theta_0)}{\cos^3(\theta_0)}, \label{r56expression} \\ 
R_{65} & = & -\frac{eV}{E_0} k, \label{r65expression} \\
R_{66} & = & 1, \label{r66expression}
\end{eqnarray}
The coefficients of the second-order terms are given by:
\begin{eqnarray}
T_{555} & = & \left( \frac{eV}{E_0}k \right)^2 L_1 \frac{\theta_0 (\cos(\theta_0) - 3) -2\sin(2\theta_0)}{2\cos^3(\theta_0)}, \label{t555expression} \\
T_{556} & = &  - \frac{eV}{E_0}k  L_1 \frac{\theta_0 (\cos(\theta_0) - 3) -2\sin(2\theta_0)}{\cos^3(\theta_0)}, \label{t556expression} \\ 
T_{566} & = & L_1 \frac{\theta_0 (\cos(\theta_0) - 3) -2\sin(2\theta_0)}{2\cos^3(\theta_0)}, \label{t566expression} \\ 
T_{655} & = & T_{656} = T_{666} = 0. 
\end{eqnarray}

\subsection{Linear analysis and initial choice of parameters}

Having constructed the transfer map, the next step is to determine appropriate
values for the parameters of the main components, based on consideration of the
linear dynamics.  The first consideration is the compression factor that the bunch
compressor should achieve.  This can be written in terms of the mean square values
of the initial and final longitudinal co-ordinates:
\begin{equation}
\frac{\langle z_2^2 \rangle}{\langle z_0^2 \rangle} = \frac{1}{b^2},
\end{equation}
where the brackets $\langle \cdot \rangle$ indicate an average over all particles in the
bunch, and $b$ is the required compression factor.  If we assume that the beam at the entrance
to the bunch compressor has no initial chirp, so that $\langle z_0 \delta_0 \rangle = 0$,
then taking only the linear terms from Eq.~(\ref{bcfullmapz}) gives:
\begin{equation}
\langle z_2^2 \rangle = R_{55}^2 \langle z_0^2 \rangle + R_{56}^2 \langle \delta_0^2 \rangle = \frac{\langle z_0^2 \rangle}{b^2}.
\label{bcconstraint1}
\end{equation}
Generally, the beam exiting the bunch compressor should have no energy chirp, so that
$\langle z_2 \delta_2 \rangle = 0$.  Using Eqs.~(\ref{bcfullmapz}) and (\ref{bcfullmapdelta})
and again keeping only linear terms gives:
\begin{equation}
\langle z_2 \delta_2 \rangle = R_{55}R_{65}\langle z_0^2 \rangle + R_{56} \langle \delta_0^2 \rangle = 0.
\label{bcconstraint2}
\end{equation}
Note that we have used the fact that $R_{66} = 1$.  A third (and final) constraint comes
from the fact that the linear transfer map must be symplectic: this means that the transfer
matrix (with elements $R_{ij}$) must have unit determinant:
\begin{equation}
R_{55} - R_{56} R_{65} = 1.
\label{bcconstraint3}
\end{equation}
With given initial bunch length, energy spread, and specified compression factor,
Eqs.~(\ref{bcconstraint1}), (\ref{bcconstraint2}) and (\ref{bcconstraint3}) can be
solved to give values for $R_{55}$, $R_{56}$ and $R_{65}$.  The result is:
\begin{eqnarray}
R_{55} & = & \frac{1}{b^2}, \\
R_{56} & = & \pm \frac{\sqrt{b^2 - 1}}{b^2} \sqrt{\frac{\langle z_0^2 \rangle}{\langle \delta_0^2 \rangle}}, \\
R_{65} & = & \mp \sqrt{b^2 - 1} \sqrt{\frac{\langle \delta_0^2 \rangle}{\langle z_0^2 \rangle}}.
\end{eqnarray}

\begin{table}
\caption{Design specifications for the bunch compressor in the International Linear Collider.
\label{ilcbunchcompressorspecifications}}
\begin{center}
\begin{tabular}{lcc}
\hline
Initial rms bunch length  & $\sqrt{\langle z_0^2 \rangle}$ & 6 mm \\
Initial rms energy spread & $\sqrt{\langle \delta_0^2 \rangle}$ & 0.15\% \\
Final rms bunch length    & $\sqrt{\langle z_2^2 \rangle}$ & 0.3 mm \\
\hline
\end{tabular}
\end{center}
\end{table}

\begin{figure}
\begin{center}
\includegraphics[width=0.32\textwidth]{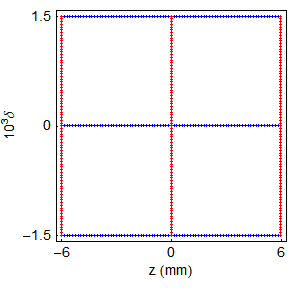}
\includegraphics[width=0.32\textwidth]{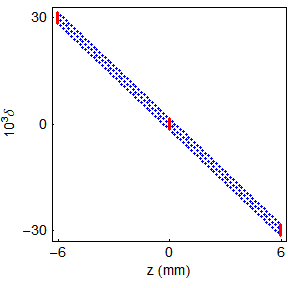}
\includegraphics[width=0.32\textwidth]{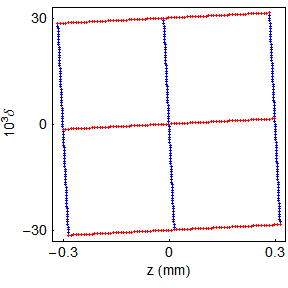}
\end{center}
\caption{Effect of the linear part of the map for a bunch compressor on particles
in longitudinal phase space.  Left: initial ``window frame'' distribution.  Middle:
distribution after passing through the RF cavity, which imparts a chirp (correlation
between energy and longitudinal position) to the bunch.  Right: final distribution
after exiting the chicane.
\label{bunchcompressorphasespacetransformationslinear}}
\end{figure}

As an example, let us consider the parameters for the bunch compressors in the
International Linear Collider, given in Table~\ref{ilcbunchcompressorspecifications}.
With the given initial bunch length and energy spread, and the specified final bunch length
(and also since $R_{56}$ must be positive in this case), we find that:
\begin{eqnarray}
R_{55} & = & 0.0025, \label{linearbcr55} \\
R_{56} & = &  0.19975\,\textrm{m}, \label{linearbcr56} \\
R_{65} & = & -4.9937\,\textrm{m}^{-1}. \label{linearbcr65} 
\end{eqnarray}
Using these values, we can illustrate the effect of the linearised bunch
compressor map on phase space.  For clarity and convenience, we use an
artificial ``window frame'' distribution: see
Fig.~\ref{bunchcompressorphasespacetransformationslinear}.
Particles are initially distributed along lines in phase space corresponding to
one standard deviation of the nominal distribution.  We see that after passing
through the bunch compressor, and applying only the linear terms in the transfer
map, the rms bunch length is reduced by a factor of 20 as required.  The rms
energy spread is \emph{increased} by the same factor, but this is unavoidable,
because the transfer map is symplectic; this means that phase space
areas are conserved under the transformation, i.e.~the longitudinal emittance
is conserved (as required by Liouville's theorem).

Based on the values for the coefficients in the transfer map given in (\ref{linearbcr55}),
(\ref{linearbcr56}) and (\ref{linearbcr65}) we can select appropriate
values for the parameters of the main components in the bunch compressor.  These will
also be determined by technical considerations and constraints from the overall design
of the facility.  Such considerations are beyond the scope of this discussion; but it is
found that the parameters given in Table~\ref{ilcbunchcompressorparameters} are
suitable for the case of the International Linear Collider.

\begin{table}
\caption{Parameters of components in a bunch compressor designed (on the basis
of the linear part of the transfer map) to achieve the specifications given in
Table~\ref{ilcbunchcompressorspecifications}.
\label{ilcbunchcompressorparameters}}
\begin{center}
\begin{tabular}{lcc}
\hline
Beam energy  & $E_0$           & 5.00 GeV     \\
RF frequency             & $f$ & 1.3 GHz   \\
RF voltage               & $V$ & 916 MV    \\
Dipole spacing           & $L_1$           & 36.3 m    \\
Dipole bending angle     & $\theta_0$      & 3.00$^\circ$ \\
\hline
\end{tabular}
\end{center}
\end{table}

\subsection{Nonlinear analysis and optimisation of parameters}

\begin{figure}
\begin{center}
\includegraphics[width=0.32\textwidth]{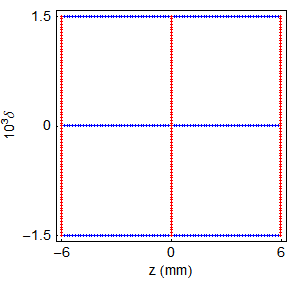}
\includegraphics[width=0.32\textwidth]{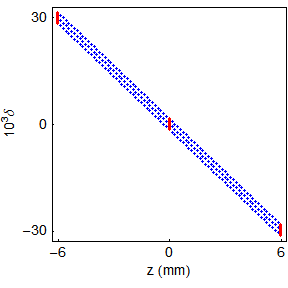}
\includegraphics[width=0.32\textwidth]{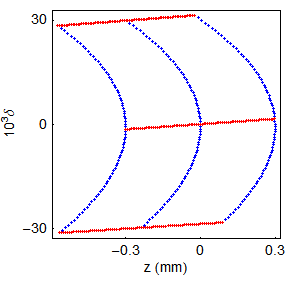}
\end{center}
\caption{Effect of the full nonlinear map for a bunch compressor on particles
in longitudinal phase space.  Left: initial ``window frame'' distribution.  Middle:
distribution after passing through the RF cavity, which imparts a chirp (correlation
between energy and longitudinal position) to the bunch.  Right: final distribution
after exiting the chicane.
\label{bunchcompressorphasespacetransformationsnonlinear}}
\end{figure}

Using the values of the parameters given in Table~\ref{ilcbunchcompressorparameters},
we can move on to the next step in the design process,
which is to evaluate the impact of nonlinear effects.  As before, we illustrate the effect
of the bunch compressor map on phase space using a ``window frame'' distribution;
this time, however, we can apply the full transfer map expressed in Eqs.~(\ref{rfmap1}),
(\ref{rfmap2}), (\ref{chicanetransfermapdz}) and (\ref{chicanetransfermapddelta}).
The results are shown in  Fig.~\ref{bunchcompressorphasespacetransformationsnonlinear}.
Although the bunch length has been reduced, there is significant distortion of the distribution:
the rms bunch length will be significantly longer than the specification requires.
To reduce the distortion, we first need to understand where it comes from,
which means looking at the map more closely.  Consider a particle entering the bunch
compressor with initial energy deviation $\delta_0 = 0$.  From the shape of the final
phase space\footnote{Note that points along lines of constant energy deviation in the initial
phase space are coloured blue: because of the rotation of phase space, the $\delta$ axis
in the final phase space plot corresponds (approximately) to the $z$ axis in the initial
phase space plot.}, we see that the final co-ordinate $z_2$ varies \emph{quadratically} with
the initial $z_0$:
\begin{equation}
z_2 \propto z_0^2.
\end{equation}
This suggests that the distortion is dominated by the term in the transfer map with
coefficient $T_{555}$.  To eliminate the distortion, we see from (\ref{t555expression})
that we would need to choose a bending angle $\theta_0$ in the chicane that satisfies:
\begin{equation}
\theta_0 (\cos(\theta_0) - 3) - 2 \sin(\theta_0) = 0.
\end{equation}
Unfortunately, the only solution to this equation is $\theta_0 = 0$, which would eliminate
the chicane altogether, and is incompatible with the linear part of the transfer map.
To address the problem of the phase space distortion caused by the nonlinearities,
we need to introduce an additional degree of freedom.  

Understanding the physical
origin of the damaging nonlinearity will help us to identify a possible solution.  Since the
RF cavity has no effect on the longitudinal co-ordinate, the $T_{555}$ term in the
transfer map must come from the chicane: this is confirmed by our observation above,
that $T_{555} = 0$ if $\theta_0 = 0$, i.e.~if we eliminate the chicane altogether.
The physical cause of the nonlinear distortion is the second-order dependence of
the change in longitudinal co-ordinate on the energy deviation for a particle passing through
a dipole.  Unfortunately, this effect is intrinsic to dipole magnets, and cannot be
corrected within the magnet itself.  However, we can compensate for it if we introduce
a second-order dependence of the energy deviation on the longitudinal co-ordinate before
the bunch enters the chicane.

To see how this works, suppose that we write the transformation of the energy deviation
in the first part of the bunch compressor as follows:
\begin{equation}
\delta_1 = R_{65}^{(1)} z_0 + T_{655}^{(1)} z_0^2. \label{bcsimplified1}
\end{equation}
We use the superscript on $R_{ij}^{(1)}$ and $T_{ijk}^{(1)}$ to indicate that
the coefficient is for the map for
only the first part of the bunch compressor.  Then, we write the transformation for the
longitudinal co-ordinate in the second part of the bunch compressor:
\begin{equation}
z_2 = R_{56}^{(2)} \delta_1 + T_{566}^{(2)} \delta_1^2. \label{bcsimplified2}
\end{equation}
Substituting $\delta_1$ from (\ref{bcsimplified1}) into (\ref{bcsimplified2}), and
keeping terms only up to second order, we find:
\begin{equation}
z_2 = R_{56}^{(2)} R_{65}^{(1)} z_0 + R_{56}^{(2)} T_{655}^{(1)} z_0^2 +
T_{566}^{(2)} \left(R_{65}^{(1)}\right)^2 z_0^2.
\end{equation}
In the transfer map we have used so far for the RF cavity in the bunch compressor,
$T_{655}^{(1)} = 0$: there is then a second-order dependence of $z_2$ on $z_0$,
with coefficient $T_{566}^{(2)}\left( R_{65}^{(1)}\right)^2$.  To eliminate this
dependence, we need to modify the transformation in the RF cavity so that:
\begin{equation}
T_{555} = R_{56}^{(2)} T_{655}^{(1)} + T_{566}^{(2)} \left(R_{65}^{(1)}\right)^2 = 0.
\label{nonlineardistortioncorrection}
\end{equation}
To achieve this, we simply need to change the phase of the RF cavity, so that
a particle at the centre of the bunch sees a non-zero accelerating (or decelerating) field:
\begin{equation}
\delta_{1a} = \delta_0 - \frac{eV}{E_0} \sin(kz_0 + \phi_0).
\end{equation}
Although this will introduce the terms that we need in the map to correct the nonlinear
distortion, it also means that there is an overall change in the bunch energy.  The overall
change in energy is inconvenient for the description of the dynamics in the chicane:
in order to maintain good accuracy in the power series representation of the transfer map,
it is desirable for particles in the bunch to be described by small values of the energy
deviation, with mean of zero.  To ensure that this is the case, we change the reference
energy from $E_0$ to $E_0 - eV\sin(\phi_0)$.  This change in reference energy will restore
the mean energy deviation of the bunch to zero; but we also need to take into account, from Eq.~(\ref{energydeviationdefinition}), that when we change the reference energy from
$E_0$ to a new value $E_0^\prime$, the energy deviation changes from $\delta$ to $\delta^\prime$:
\begin{equation}
\delta^\prime = \left( \delta - \frac{\Delta E_0}{E_0} \right) \frac{E_0}{E_0^\prime},
\end{equation}
where $\Delta E_0 = E_0^\prime - E_0$.  This scaling of the momentum deviation,
with no associated change in the longitudinal co-ordinate, is a non-symplectic transformation,
and leads to a change in the longitudinal emittance.  A similar effect occurs in the transverse
degrees of freedom: when associated with acceleration in a linac, it leads to a reduction
of the emittances in all three planes, and is known as \emph{adiabatic damping}.
In the present case of the bunch compressor, the result is that the full transformation in
the RF cavity is now:
\begin{eqnarray}
z_1 & = & z_0, \\
\delta_1 & = & \left( \delta_0 - \frac{eV}{E_0} ( \sin(kz_0 + \phi_0) - \sin(\phi_0) ) \right)
\left( 1 - \frac{eV}{E_0} \sin(\phi_0) \right)^{\!-1}.
\end{eqnarray}

Expanding to second order in $z_0$, this gives:
\begin{equation}
\delta_1 = \left( 1 - \frac{eV}{E_0} \sin(\phi_0) \right)^{\!-1}
\left( \delta_0 - \frac{eV}{E_0} \cos(\phi_0) k z_0
+ \frac{eV}{2E_0} \sin(\phi_0) k^2 z_0^2 + \ldots \right) 
\end{equation}
Hence:
\begin{equation}
T_{655}^{(1)} = \frac{eV}{2E_0} \sin(\phi_0) k^2.
\end{equation}
A non-zero value of $\phi_0$ leads to a non-zero value for $T_{655}^{(1)}$, and allows
us to satisfy the requirement expressed in Eq.~(\ref{nonlineardistortioncorrection}),
to correct the nonlinear distortion arising from $T_{566}^{(2)}$.  The other constraints
that we need to satisfy are, as before, the specified reduction in bunch length,
Eq.~(\ref{bcconstraint1}):
\begin{equation}
R_{55}^2 + \frac{\langle \delta_0^2 \rangle}{\langle z_0^2 \rangle} R_{56}^2 = \frac{1}{b^2},
\label{bcconstraint1a}
\end{equation}
and the requirement for zero final energy chirp, Eq.~(\ref{bcconstraint2}):
\begin{equation}
R_{55}R_{65} + \frac{\langle \delta_0^2 \rangle}{\langle z_0^2 \rangle} R_{56}R_{66} = 0.
\label{bcconstraint2a}
\end{equation}
Note that the expressions for the coefficients $R_{ij}$ need to be re-calculated for the case
$\phi_0 \neq 0$: in particular $R_{66}$ is no longer equal to 1.  Also, the constraint
expressed in Eq.~(\ref{bcconstraint2}) no longer applies, because of the (non-symplectic)
rescaling of the reference energy that we have applied following the RF cavity.

We therefore have three constraints, Eq.~(\ref{nonlineardistortioncorrection}),
(\ref{bcconstraint1a}) and (\ref{bcconstraint2a}).  Assuming fixed RF frequency and
dipole spacing\footnote{The RF frequency and dipole spacing are fixed at time of
machine construction; the RF voltage, RF phase, and dipole bending angle are readily adjusted
during operation.}, we can satisfy the constraints by adjusting the values of the RF voltage
$V$, the RF phase $\phi_0$, and the dipole bending angle $\theta_0$.  Given the complicated,
nonlinear nature of the constraint equations, the parameter optimisation is best done
numerically.  Suitable values for the parameters are shown in
Table~\ref{ilcbunchcompressorparametersnonlinear}.

\begin{table}
\caption{Parameters of components in a bunch compressor designed to achieve the
specifications given in Table~\ref{ilcbunchcompressorspecifications}, and taking into
account the effects of nonlinear dynamics.  The parameter values are to be compared
with those in Table~\ref{ilcbunchcompressorparameters}, which were determined on
the basis only of the linear dynamics.
\label{ilcbunchcompressorparametersnonlinear}}
\begin{center}
\begin{tabular}{lcc}
\hline
Initial beam energy  & $E_0$           & 5.00 GeV     \\
Final beam energy  & $E_0^\prime$           & 4.43 GeV     \\
RF frequency             & $f$ & 1.3 GHz   \\
RF voltage               & $V$ & 1079 MV    \\
RF phase               & $\phi_0$ & 31.86$^\circ$    \\
Dipole spacing           & $L_1$           & 36.3 m    \\
Dipole bending angle     & $\theta_0$      & 2.83$^\circ$ \\
\hline
\end{tabular}
\end{center}
\end{table}

\begin{figure}
\begin{center}
\includegraphics[width=0.32\textwidth]{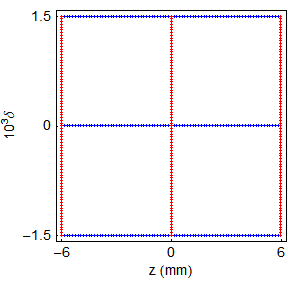}
\includegraphics[width=0.32\textwidth]{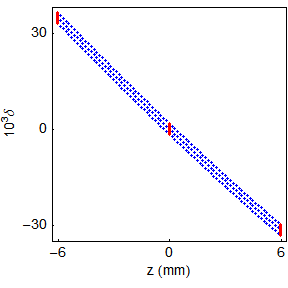}
\includegraphics[width=0.32\textwidth]{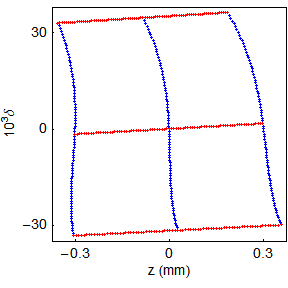}
\end{center}
\caption{Effect of the full nonlinear map for a bunch compressor on particles
in longitudinal phase space, with parameters adjusted to compensate for the
nonlinear distortion seen in Fig.~\ref{bunchcompressorphasespacetransformationsnonlinear}.
Left: initial ``window frame'' distribution.  Middle:
distribution after passing through the RF cavity, which imparts a chirp (correlation
between energy and longitudinal position) to the bunch.  Right: final distribution
after exiting the chicane.
\label{bunchcompressorphasespacetransformationsnonlinearcompensated}}
\end{figure}

As before, we illustrate the effect of the bunch compressor on phase space
using a ``window frame'' distribution.  But now we use the parameters given
in Table~\ref{ilcbunchcompressorparametersnonlinear} to aim to compress
the bunch length by a factor 20 while minimising the second-order distortion: the
results are shown in Fig.~\ref{bunchcompressorphasespacetransformationsnonlinearcompensated}.
We see that the nonlinear distortion is greatly reduced: the remaining distortion
now appears to be dominated by third-order terms, and appears to be
small enough that it should not significantly affect the performance of the machine.

\subsection{Summary: lessons from analysis of the bunch compressor}

The analysis of the bunch compressor presented in this section highlights some
important aspects of nonlinear dynamics in accelerators.  First, we see that
nonlinear effects can limit the performance of an accelerator system.  Sometimes
the effects are small enough that they can be ignored; however, in many cases, a system
designed without taking account of nonlinearities will not achieve the specified performance.
Secondly, a careful analysis can lead to an understanding of the nonlinear behaviour of
a system, including the origin of the nonlinearities.  Based on this analysis, it may be
possible to find a means of compensating any adverse effects.

\section{Nonlinear dynamics in a storage ring}

In this section, we consider some of the phenomena associated with nonlinear dynamics
in periodic beamlines, using a storage ring as an example.  We will explain the significance
of \emph{symplectic maps}, and describe some of the challenges in constructing and
applying symplectic maps.  Finally, we will outline some of the analysis methods that can
be used to characterise nonlinear beam dynamics in periodic beamlines.

In the discussion of a bunch compressor in the previous section of these notes, we focused
on the longitudinal dynamics.  For the case of a storage ring, we shall be concerned with
the transverse dynamics.  Nonlinear effects can impact both longitudinal and transverse
motion, and in many cases it may be necessary to consider both.  This is indeed the case in
many storage rings; but for simplicity, we shall restrict the discussion to the transverse
dynamics.

We shall make the following assumptions:
\begin{itemize}
\item the storage ring is constructed from some number of identical cells
consisting of dipoles, quadrupoles and sextupoles;
\item the phase advance per cell can be tuned from close to zero, up to about
0.5$\times 2\pi$.
\item there is a single sextupole per cell, located at a point where the
horizontal beta function is $1\,$m, and the alpha function is zero.
\end{itemize}
Usually, storage rings will contain (at least) two sextupoles per cell, to correct horizontal
and vertical chromaticity.  Again for the sake of simplicity, we will use only one sextupole
per cell.

\subsection{Correction of chromaticity using sextupole magnets}

\begin{figure}
\begin{center}
\includegraphics[width=0.9\textwidth]{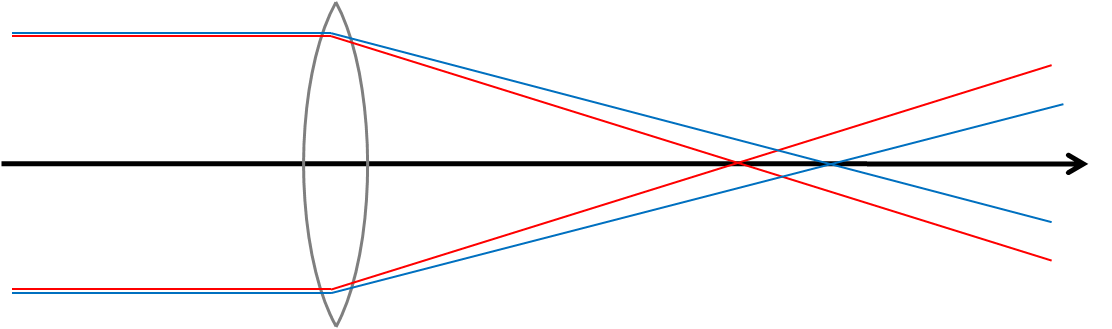}
\end{center}
\caption{Chromaticity in a quadrupole magnet.  The beam rigidity increases
with energy, leading to an increase in the focal length of a quadrupole with
particle energy.
\label{figchromaticity}}
\end{figure}

Sextupoles are needed in a storage ring to compensate for the fact that
quadrupoles have lower focusing strength for particles of higher energy
(see Fig.~\ref{figchromaticity}): this is characterised by the chromaticity, which is
defined as the change in particle tune in a storage ring with respect to energy deviation.
Chromaticity often has undesirable consequences.  For example, if particles with
sufficiently large energy deviation (by virtue of the natural energy spread of the beam)
have betatron tunes close to integer or half-integer values, then
the motion of these particles can become unstable, leading to loss of the particles
from the storage ring.

\begin{figure}
\begin{center}
\includegraphics[width=0.75\textwidth]{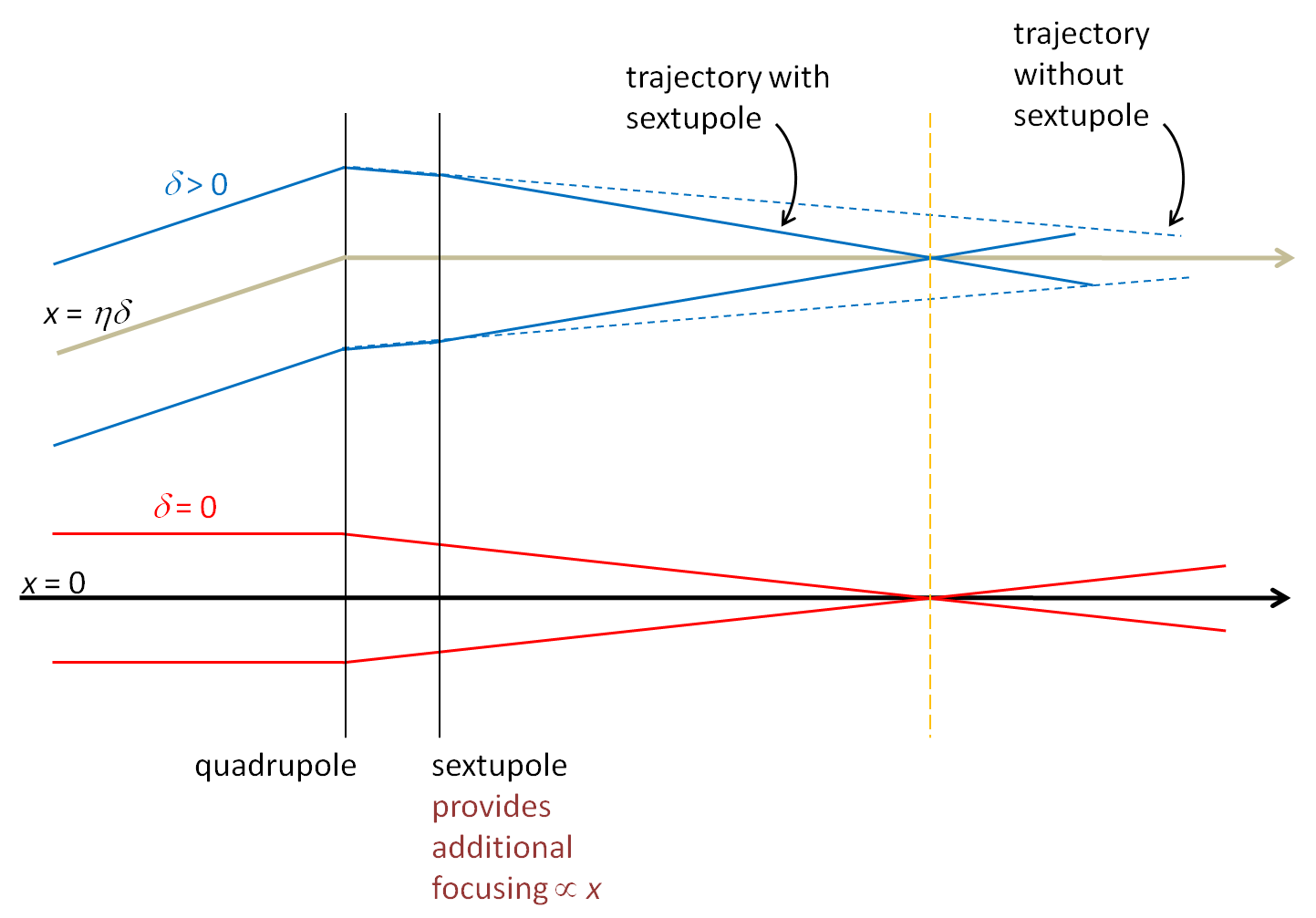}
\end{center}
\caption{Correction of chromaticity in a quadrupole magnet, using a sextupole at a
location with non-zero dispersion.
\label{chromaticcorrection}}
\end{figure}

A sextupole can be regarded as a quadrupole with focusing strength that
increases with horizontal (transverse) distance from the axis of the magnet.
If sextupoles are located where there is non-zero dispersion, they can be used
to control the chromaticity in a storage ring.  The dispersion describes the change
in position of the closed orbit with respect to changes in energy deviation: particles
with higher energy deviation will then pass through the sextupole further from the
axis of the magnet (see Fig.~\ref{chromaticcorrection}), and hence receive
additional focusing or defocusing.  By adjusting the strength of the sextupole (depending
on the size of the dispersion), it is possible to compensate the natural
chromaticity arising from the variation of the focal lengths of the quadrupoles with
particle energy.  

The chromaticity of a storage ring, and hence the strengths of the sextupoles needed
to correct the chromaticity, will normally be a function of the phase advance per cell.
However, to investigate and illustrate the nonlinear effects of the sextupoles, we shall
assume that we keep the strength $k_2L$  of the sextupoles in our simple storage ring
fixed when we change the phase advance.  Although this means that the chromaticity
will in general be non-zero, since we shall be looking only at the motion of particles with
zero energy deviation, the chromaticity will play no real role.  The effects of the sextupoles
in this case are sometimes called the \emph{geometric} effects, to distinguish them from
the chromatic effects.  The geometric effects  of sextupoles can have significant adverse
impact on the stability of the motion of particles in a storage ring.  Particularly for low-emittance
electron storage rings, for example in third-generation synchrotron light sources, it is an
important task at the design stage to optimize the chromatic effects of the sextupoles while
minimising the geometric effects, in order to achieve a good beam lifetime.

In our simple storage ring, we can assume that the map from one sextupole to the next
is linear, and corresponds to a rotation in phase space through an angle equal to the phase
advance:
\begin{equation}
\left(
\begin{array}{c}
x \\
p_x
\end{array}
\right) \mapsto
\left(
\begin{array}{cc}
\cos (\mu_x) & \sin (\mu_x) \\
-\sin (\mu_x) & \cos (\mu_x) \\
\end{array}
\right)
\left(
\begin{array}{c}
x \\
p_x
\end{array}
\right). \label{linearmap1}
\end{equation}
Note that we use the symbol $\mapsto$ to mean ``transforms to'': this avoids the need
to use subscripts to indicate initial and final values of the phase space variables (as in,
for example, $x_0$ and $x_1$).
Again to keep things simple, we shall consider only horizontal motion, and
assume that the vertical co-ordinate $y=0$.

The change in the horizontal momentum of a particle moving through the sextupole
is found by integrating the Lorentz force over the length $L$ of the sextupole:
\begin{equation}
\Delta p_x = -\int_0^L \frac{B_y}{B\rho}\, ds.
\end{equation}
The sextupole strength $k_2$ is defined by:
\begin{equation}
k_2 = \frac{1}{B\rho} \frac{\partial^2 B_y}{\partial x^2},
\end{equation}
where $B\rho$ is the beam rigidity.  For a pure sextupole field, in the case that
the vertical co-ordinate $y=0$, the vertical component of the magnetic field is given by:
\begin{equation}
\frac{B_y}{B\rho} = \frac{1}{2} k_2 x^2.
\end{equation}
If the sextupole is short (compared to the betatron wavelength), then we can neglect
the small change in the co-ordinate $x$ as the particle moves through the sextupole,
in which case:
\begin{equation}
\Delta p_x \approx -\frac{1}{2} k_2 L x^2.
\end{equation}
Hence, the transfer map for a particle moving through a short sextupole can be represented by a
``kick'' in the horizontal momentum:
\begin{eqnarray}
x & \mapsto & x, \label{sextupolekick1} \\
p_x & \mapsto & p_x - \frac{1}{2} k_2 L x^2. \label{sextupolekick2}
\end{eqnarray}

\subsection{Effects of sextupoles in a storage ring: dependence on phase advance}

\begin{figure}
\begin{center}
\begin{tabular}{cc}
$\mu_x = 0.202\times 2\pi$ & $\mu_x = 0.252\times 2\pi$ \\
\includegraphics[width=0.4\textwidth]{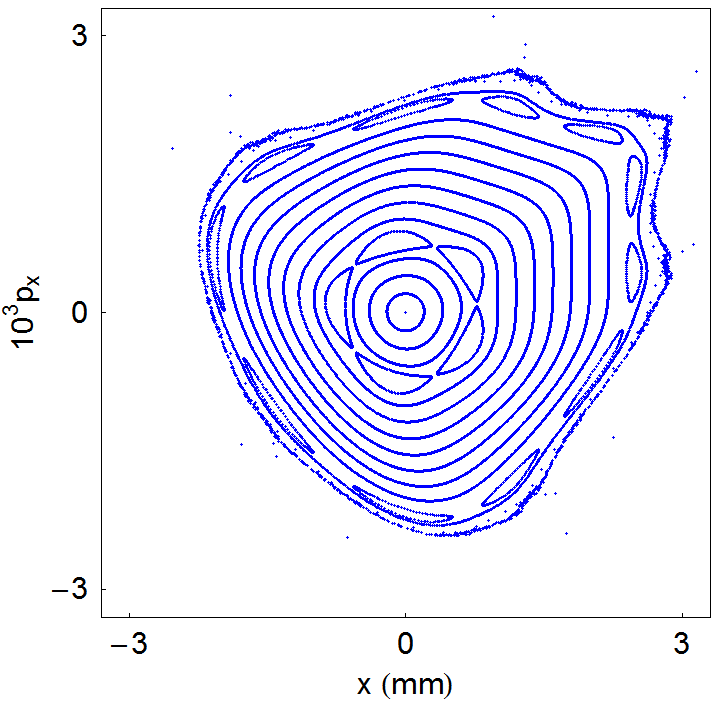} &
\includegraphics[width=0.4\textwidth]{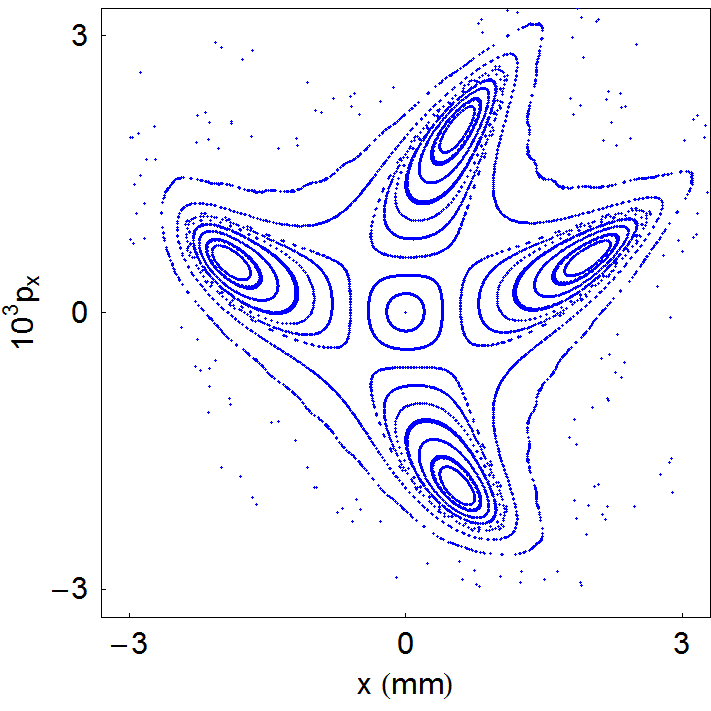} \\
$\mu_x = 0.330\times 2\pi$ & $\mu_x = 0.402\times 2\pi$ \\
\includegraphics[width=0.4\textwidth]{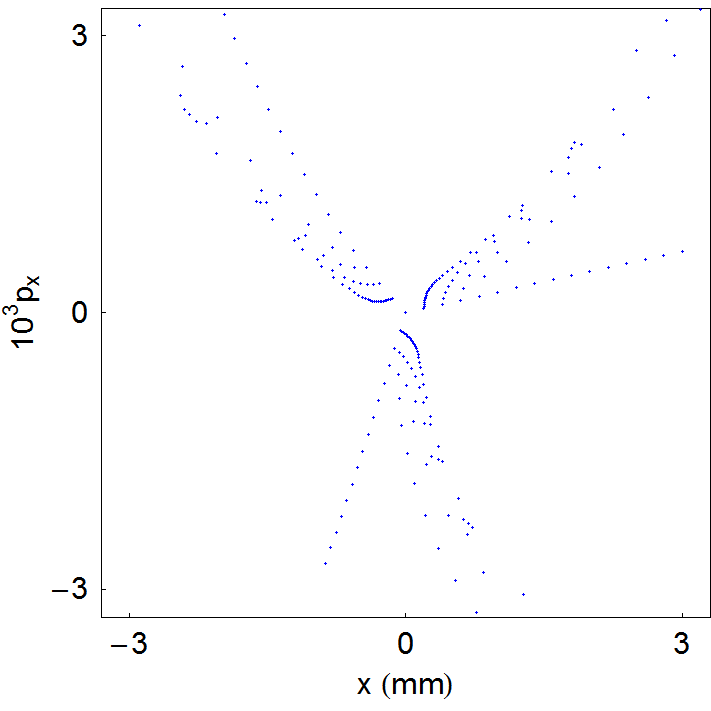} &
\includegraphics[width=0.4\textwidth]{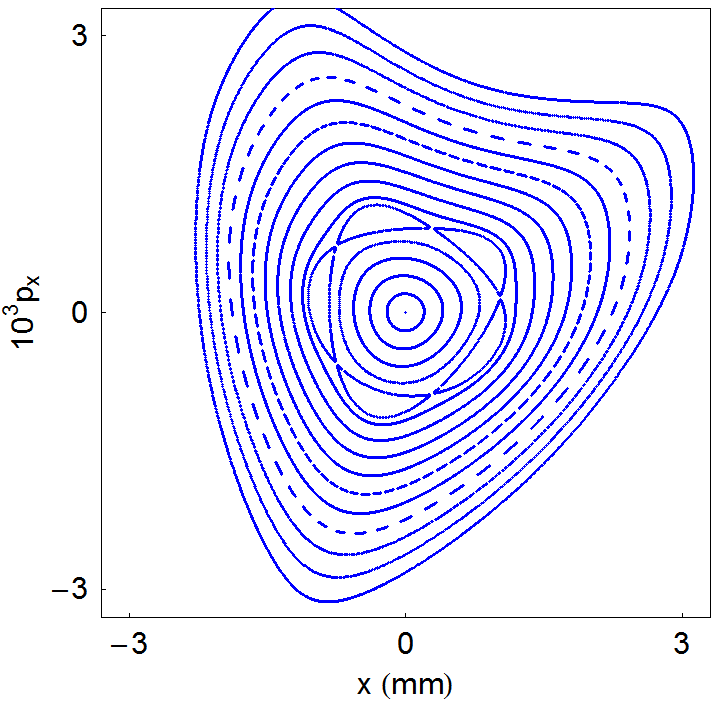} \\
\multicolumn{2}{c}{$\mu_x = 0.490\times 2\pi$} \\
\multicolumn{2}{c}{\includegraphics[width=0.4\textwidth]{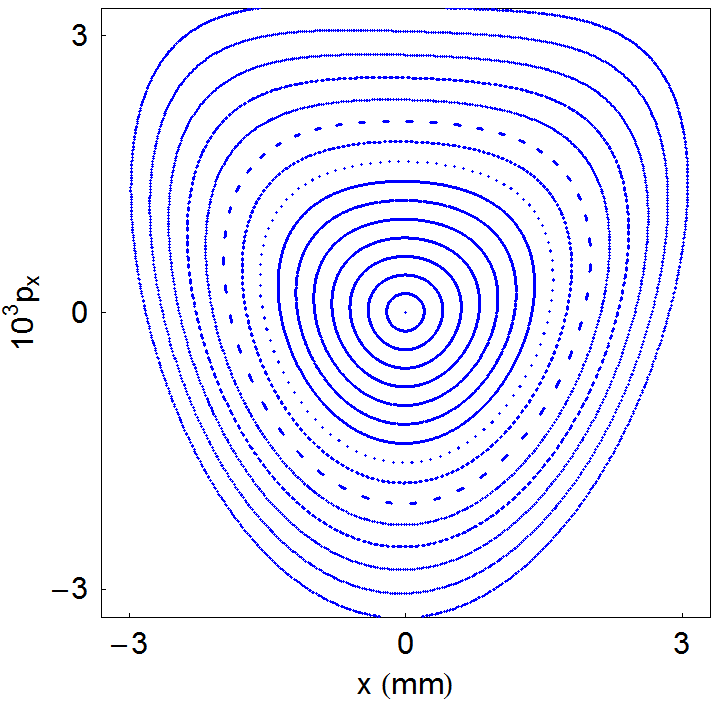} \label{ps5}} \\
\end{tabular}
\end{center}
\caption{Phase space portraits for a simple storage ring, tuned to various
different values of the phase advance per cell, and containing (in addition
to the quadrupoles) a single sextupole per cell.
\label{ps1}}
\end{figure}

To illustrate the effect of the sextupole in our storage ring, let us choose a fixed
value $k_2L=-600\,\textrm{m}^{-2}$ for the strength of the sextupole, and look
at the effects of the maps for different phase advances.  As mentioned above, we
shall only consider the case that the particles have zero energy deviation.  For a
given value of the phase advance, we construct a \emph{phase space portrait} by
plotting the values of the dynamical variables after repeated application of the
transfer map (equation (\ref{linearmap1}), followed by (\ref{sextupolekick1})
and (\ref{sextupolekick2})) for a range of initial conditions. The phase space
portraits for the simple storage ring with a range of phase advances from
$0.2\times 2\pi$ to $0.5\times 2\pi$ are shown in Fig.~\ref{ps1}.

There are some interesting features in these phase space portraits to which it is
worth drawing attention:
\begin{itemize}
\item For small amplitudes (small $x$ and $p_x$), particles trace out closed loops
around the origin: this is what we expect for a purely linear map.
\item As the amplitude is increased, ``islands'' appear in the phase space, with the
number of islands related to the phase advance:
the phase advance (for the linear map) is often close to $m/p$ where
$m$ is an integer and $p$ is the number of islands.
\item Sometimes, a larger number of islands appears at larger amplitude.
\item Usually, there is a closed curve that divides a region of stable motion
from a region of unstable motion.  Outside that curve, the amplitude of
particles increases without limit as the map is repeatedly applied, i.e.~the motion
is unstable.
\item The area of the stable region depends strongly on the phase advance:
for a phase advance close to $2\pi /3$, it appears that the stable region
almost vanishes altogether.
\item As the phase advance is increased towards $\pi$, the
stable area becomes large, and distortions from the linear ellipse become
less evident.
\end{itemize}

An important observation is that the effect of the sextupole in the periodic cell depends
strongly on the phase advance across the cell.
We can start to understand the significance of the phase advance by considering two special cases:
first, the case when the phase advance is equal to an integer times $2\pi$;
and second, when the phase advance is equal to a half integer times $2\pi$.
In the first case, when the phase advance is an integer, the linear part of the map is just the identity:
\begin{eqnarray}
x & \mapsto & x, \\
p_x & \mapsto & p_x.
\end{eqnarray}
So the combined effect of the linear map and the sextupole kick is:
\begin{eqnarray}
x & \mapsto & x, \\
p_x & \mapsto & p_x - \frac{1}{2}k_2 L x^2.
\end{eqnarray}
Clearly, for $x \neq 0$, the horizontal momentum will increase without limit.
There are no stable regions of phase space, apart from the line $x=0$.

Now consider what happens in the second case, when the phase advance over one cell
is a half integer times $2\pi$, so the linear part of the map is just a rotation by $\pi$.
If a particle starts at the entrance of a sextupole with $x = x_0$ and $p_x = p_{x0}$,
then at the exit of that sextupole (using the subscript notation to indicate initial and final
values of the variables):
\begin{eqnarray}
x_1 & = & x_0, \\
p_{x1} & = & p_{x0} - \frac{1}{2}k_2 L x_0^2.
\end{eqnarray}
Then, after passing to the entrance of the next sextupole, the phase space variables will be:
\begin{eqnarray}
x_2 & = & -x_1 = -x_0, \\
p_{x2} & = & - p_{x1} = - p_{x0} + \frac{1}{2}k_2 L x_0^2.
\end{eqnarray}
Finally, on passing through the second sextupole:
\begin{eqnarray}
x_3 & = & x_2 = -x_0, \\
p_{x3} & = & p_{x2} - \frac{1}{2}k_2 L x_2^2 = - p_{x0}.
\end{eqnarray}
In other words, the momentum kicks from the two sextupoles in successive cells
cancel each other exactly, and the resulting map is a purely linear phase space rotation by $\pi$.
In this situation, we expect the motion to be stable (and periodic), regardless of the amplitude.

\subsection{Resonances}

\begin{figure}
\begin{center}
\includegraphics[width=0.9\textwidth]{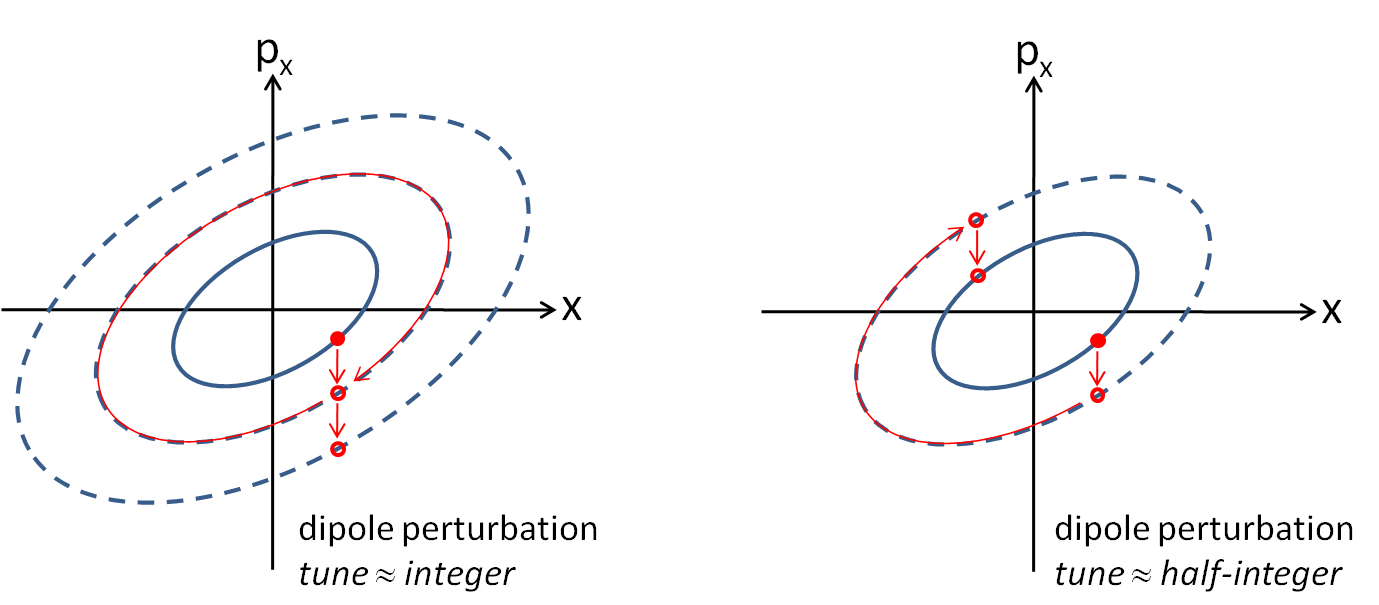} \\
\includegraphics[width=0.9\textwidth]{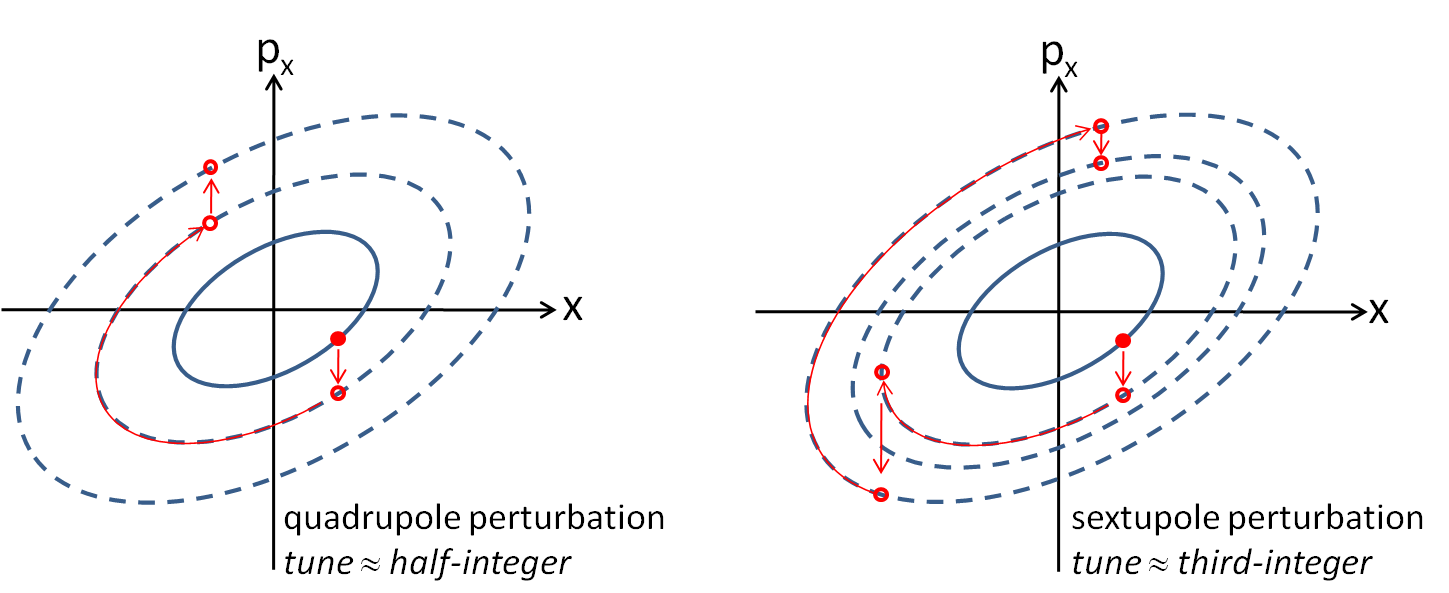}
\end{center}
\caption{Integer (top) and half-integer (bottom) resonances driven by dipole and
quadrupole perturbations (respectively) in a storage ring.
\label{resonancedriving}}
\end{figure}

\begin{figure}
\begin{center}
\includegraphics[width=0.32\textwidth]{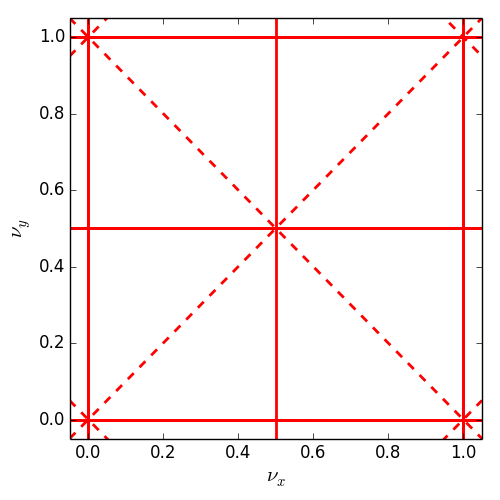}
\includegraphics[width=0.32\textwidth]{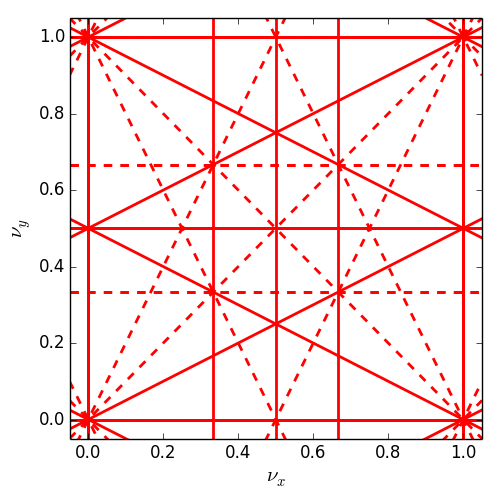}
\includegraphics[width=0.32\textwidth]{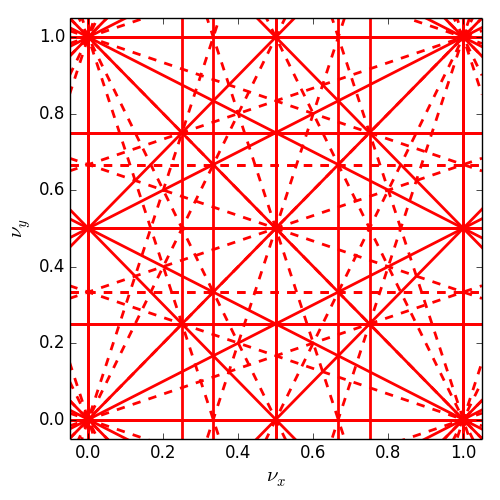}
\end{center}
\caption{Resonances in a storage ring illustrated in tune space. The horizontal
axis corresponds to the fractional part of the horizontal tune, and the vertical axis corresponds
to the fractional part of the vertical tune.  The lines show where the values of the tunes
satisfy the resonance condition $m_x \nu_x + m_y \nu_y = \ell$, for integers $m_x$, $m_y$
and $\ell$. Resonances are shown up to $2^\textrm{nd}$ order (left), $3^\textrm{rd}$ order (middle)
and $4^\textrm{th}$ order (right). Solid lines indicate resonances associated with normal multipole 
components and dashed lines indicate resonances 
associated with skew multipole components of the magnetic field expansion
\cite{wiedemann2015}. \label{resonancediagram}}
\end{figure}

The important conclusion is that the effect of sextupole ``kicks'' depends on the phase
advance between the sextupoles.  This is similar to the case of perturbations arising from
dipole and quadrupole errors in a storage ring.  In the case of dipole errors, the kicks add
up if the phase advance is an integer, and cancel if the phase advance is a half integer, see
Fig.~\ref{resonancedriving}.
In the case of quadrupole errors, the kicks add up if the phase advance is
a half integer.  In general, there are certain values of the phase advance, termed \emph{resonances},
at which small perturbations to the motion applied in each periodic section combine to
cause particle motion to be unstable.  If we include vertical as well as horizontal motion,
then we find that resonances occur when the tunes satisfy:
\begin{equation}
m_x \nu_x + m_y \nu_y = \ell,
\label{resonancecondition}
\end{equation}
where $m_x$, $m_y$ and $\ell$ are integers.  The \emph{order} of the resonance is
$|m_x| + |m_y|$: the case $|m_x| + |m_y| = 1$ is an integer resonance; 
$|m_x| + |m_y| = 2$ is a 2$^\mathrm{nd}$ order resonance (or half-integer resonance if
$m_x = 0$ or $m_y = 0$), and so on.  Resonances can be illustrated on a resonance
diagram, see Fig.~\ref{resonancediagram}.  The effect of a resonance depends on the
order of the resonance, and the presence of components or perturbations that can
``drive'' the resonance.  Although it is tempting to associate resonances of a particular
order with corresponding multipoles, in reality the situation is rather complicated: it
turns out that sextupoles (and other higher-order multipoles) can drive resonances of
many different orders depending on the exact situation.

Resonances are associated with unstable motion for particles in storage rings. However,
the number of resonance lines in tune space is infinite: any point in tune space will be close
to a resonance of some order.  This raises two questions: first, how do we know what the real
effect of any given resonance line will be?  Second, how can we design a storage ring to minimise
the adverse effects of resonances? 

From the discussion above, we have seen that for certain phase advances the kicks from sextupole 
magnetic fields can be cancelled in successive passages of a particle through a periodic cell.
It turns out that certain resonances can be suppressed by constructing a lattice with some periodicity $P$,
i.e.~building a machine from $P$ identical cells.  In that case, a resonance corresponding to
a particular value of $\ell$ in the resonance condition Eq.~(\ref{resonancecondition}) is suppressed by the lattice symmetry
(to first order) if $\ell/P$ is not an integer.
This means that the kicks from the
magnetic fields in consecutive cells cancel out over one turn, and the lattice
does not ``drive'' the resonance: the 
resonance is called \emph{non-systematic}.  Of course, this is (strictly speaking) only true if the lattice 
is perfectly periodic, and any variation, for example from random errors in the strengths of the magnets,
can break the symmetry and drive nominally non-systematic resonances.

\begin{figure}[t]
\begin{center}
\includegraphics[width=0.32\textwidth]{figures/tunediagram_periodicity1_order1-4.png}
\includegraphics[width=0.32\textwidth]{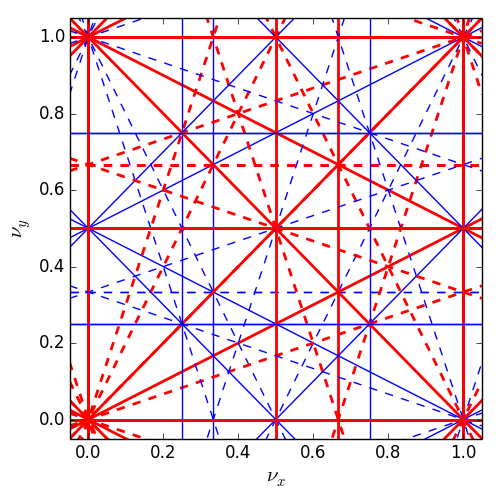}
\includegraphics[width=0.32\textwidth]{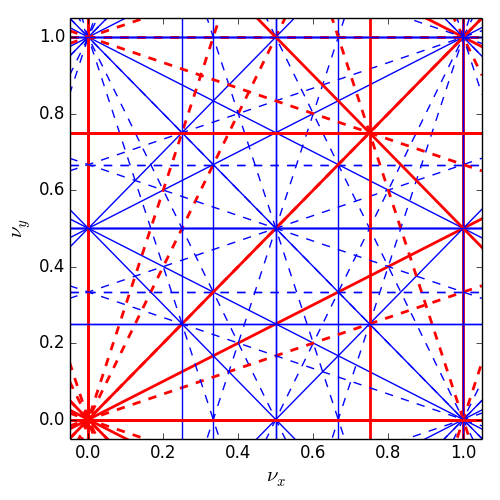}
\end{center}
\caption{Resonances up to $4^\textrm{th}$ order in a storage ring with periodicity $P=1$ (left), $P=2$ (middle)
and $P=3$ (right). Red lines indicate systematic resonances, while the blue lines show the non-systematic
resonances which (to first order) are suppressed by the lattice symmetry. Solid lines indicate 
resonances associated with normal multipole components and dashed lines indicate resonances
associated with skew multipole components of the magnetic field expansion \cite{wiedemann2015}. \label{resonancediagram2}}
\end{figure}

If $\ell/P$ (for a given resonance) is an integer, the resonance will be \emph{systematic}. In this case, the kicks from consecutive cells of the lattice add up 
coherently. Figure~\ref{resonancediagram2} shows a comparison of the resonance diagram 
for different lattice periodicities $P$.
If the
``ideal'' symmetry of a lattice is broken by random errors or other effects, then
the periodicity is effectively reduced to $P = 1$, so $\ell/P$ is an integer for
any integer $\ell$:
all resonances in that case are systematic.

One way to understand systematic resonances is to consider the ``tunes''
of individual cells in a periodic lattice (i.e.~the phase advances per cell,
divided by $2\pi$): if the periodicity is $P$, then 
each periodic cell has tunes $\nu_x/P$ and $\nu_y/P$.  It follows from the 
resonance condition Eq.~(\ref{resonancecondition}) that:
\begin{equation}
    m_x \frac{\nu_x}{P} + m_y \frac{\nu_y}{P} = \frac{\ell}{P},
\end{equation}
and hence if $\ell/P$ is an integer, the resonance condition is satisfied
for each individual cell, as well as for the full lattice.



\begin{figure}
\begin{center}
\includegraphics[trim=0 20 0 0, clip, width=0.95\textwidth]{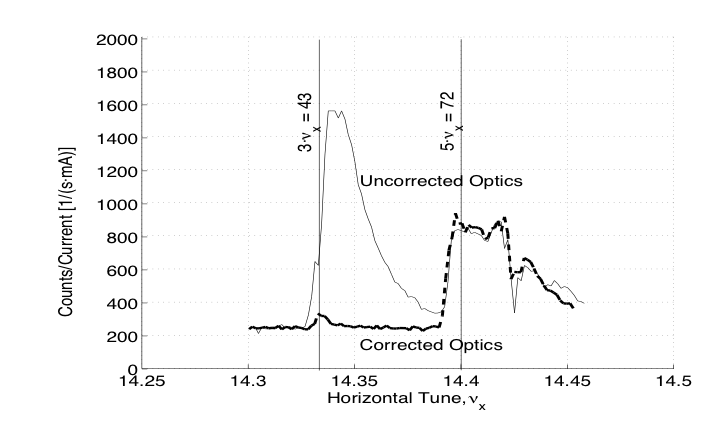} \\
\end{center}
\caption{Horizontal tune scans before (thin, solid line) and after (thick, dashed line) optics correction.  In the uncorrected optics, the $\beta$-beat is
of order 30\%; following correction, this is reduced to 1\%.
The vertical tune is kept constant at 8.15. The plot shows the count rate  measured 
in a gamma-ray detector (indicating losses of particles from the beam) divided by beam current \cite{robin2000}.
\label{latticeSymmetry}}
\end{figure}

\begin{figure}
\begin{center}
\includegraphics[width=0.77\textwidth]{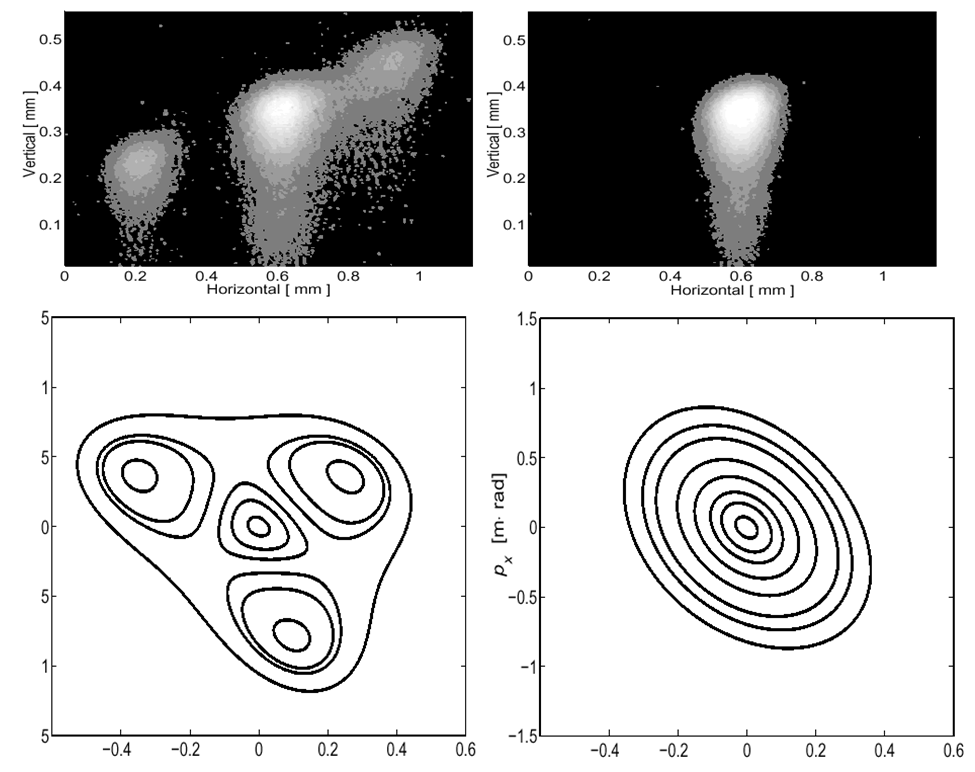} \\
\end{center}
\caption{The upper images are taken from a synchrotron radiation monitor in the Advanced Light Source
at Lawrence Berkeley National Laboratory, with the lattice tuned near the $3\nu_x$ resonance.
The image on the left was taken before correction of the optics to reduce $\beta$-beating;
the image on the right was taken after optics correction. (Note that the vertical
asymmetry is the result of some distortion in the light optics.)
The lower plots show, for comparison, corresponding horizontal phase space topologies
obtained from particle tracking in a machine model \cite{robin2000}.
\label{phasespaceislandsals}}
\end{figure}

A powerful technique to minimise the impact of resonances on the beam is therefore to design 
storage rings and synchrotrons (or the arcs of colliders) with a high periodicity. A good example
illustrating the suppression of resonances through lattice periodicity was demonstrated at the Advanced 
Light Source (ALS) at Lawrence Berkeley National Laboratory (LBNL). Figure \ref{latticeSymmetry} shows
the beam loss rate measured in the ALS for a range of different horizontal tunes in two machine configurations 
\cite{robin2000}. In the first configuration, with uncorrected linear machine optics, the $\beta$-beating 
reached approximately 30\%. In this configuration, strong losses were observed close to the 
horizontal 3$^\mathrm{rd}$ order resonance $3\nu_x=43$ even though this is a non-systematic
resonance for the ideal ALS lattice, which has periodicity $P=12$ ($\ell/P = 43/12$ is not an integer).
In the second configuration,
following correction of the linear optics (as is done in routine machine operation), the
$\beta$-beating was significantly reduced, to below the 1\% level. 
In this configuration, the beam showed only a very weak sensitivity to the resonance. Thus, by 
restoring the machine periodicity $P=12$ of the ALS by correcting the $\beta$-beating, the 
$3\nu_x=43$ resonance was almost completely suppressed. It is also interesting to take
a closer look at the loss rates observed close to the $5\nu_x = 72$ resonance, also
seen in Fig.~\ref{latticeSymmetry}. The 
level of losses at this resonance is practically the same for both machine configurations, 
i.e.~without optics correction and with optics correction. The similarity in
the losses in both configurations is explained 
by the fact that $5\nu_x = 72$ is a systematic resonance ($\ell/P = 72/12 = 6$ is an integer) 
and therefore there is no suppression of the resonance after restoring the lattice periodicity. 

Another way of demonstrating the suppression of the 3$^\mathrm{rd}$ order resonance through the lattice
periodicity was also demonstrated at the ALS, during the same experiment \cite{robin2000}. Synchrotron radiation
images were recorded for a machine working point close to the 3$^\mathrm{rd}$ order resonance 
$3\nu_x=43$ as shown in Fig.~\ref{phasespaceislandsals} (top). Before optics correction,
the image showed the beam ``split'' into several spots; after optics correction (restoring
the periodicity of the lattice) only a single beam spot is observed. This observation
can be understood from the phase space topology obtained from a machine model for
the two configurations, as shown in  
Fig.~\ref{phasespaceislandsals} (bottom). For the uncorrected optics, the resonance distorts the 
phase space and creates resonant ``islands'' in phase space.  The islands are populated by electrons 
in the beam, so that each island, in addition to the central part of the beam, acts as a source of 
synchrotron radiation. The separation between the islands in phase space (projected onto the 
co-ordinate axes) is visible on the synchrotron light monitor. This phase space structure does not exist
for the machine with restored periodicity, as the resonance is suppressed. The phase space looks
highly linear and only one beam spot is observed.
 
\begin{figure}
\begin{center}
\includegraphics[width=0.65\textwidth]{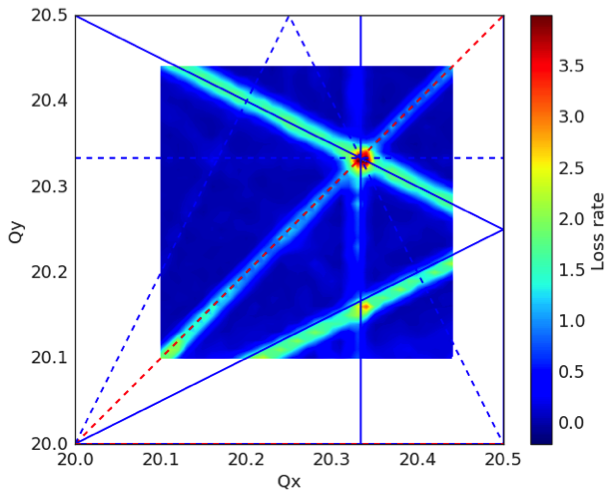} \\
\end{center}
\caption{Measurement of the beam loss rate for two-dimensional tune scan at the CERN SPS. The 
loss rate (in arbitrary units) is indicated by the color scale. The lines indicate resonances up to
$3^\textrm{rd}$ order, where systematic resonances are indicated in red and non-systematic
resonances in blue. Solid lines correspond to normal resonances, dashed lines to skew resonances.
The beam shows losses at all the non-systematic normal $3^\textrm{rd}$ order resonances, because
of linear optics distortions ($\beta$-beating).
\label{SPSlosses}}
\end{figure}

Another example of resonance excitation in operational conditions is provided by the Super
Proton Synchrotron (SPS) at CERN \cite{bartosik2011, papaphilippou2014}. Figure~\ref{SPSlosses} shows a measurement of the beam loss
rate for a two-dimensional tune scan. Even though the SPS has a design lattice periodicity of $P=6$,
which suppresses most of the $3^\textrm{rd}$ order resonances (i.e.~they are 
non-systematic), residual beam 
loss is observed on practically all $3^\textrm{rd}$ order normal resonances (solid lines). These 
resonances are observed to be excited because of perturbations of the linear optics at the level
of typically around 10\% $\beta$-beating, which in the case of the SPS cannot be corrected 
as it has no individually powered quadrupole magnets. This situation is not unusual for hadron
machines (such as the synchrotrons of the injector chain of the LHC, the Large Hadron Collider at CERN). 
If necessary, resonances are compensated by dedicated multipole corrector magnets.

The optimisation of the nonlinear beam dynamics and thus the machine performance usually
requires use of a combination of tools, such as numerical integration of particle motion (tracking)
as well as analytical tools, as we will discuss in the remaining parts of these notes.

\subsection{Symplecticity}

For any detailed analysis of nonlinear dynamics in an accelerator, we need a convenient way
to represent nonlinear transfer maps.  In our analysis of a bunch compressor in
section~\ref{bunchcompressor}, we represented the transfer maps for the RF cavity and the
chicane as power series in the dynamical variables.  For example, the longitudinal transfer map
for a chicane can be written as
\begin{eqnarray}
z_1 & = & z_0 + 2L_1 \left( \frac{1}{\cos (\theta_0)} - \frac{1}{\cos (\theta)} \right), \\
\delta_1 & = & \delta_0,
\end{eqnarray}
where 
\begin{equation}
\theta = \frac{\theta_0}{1 + \delta_0}.
\end{equation}
Expanding the map for a chicane as a power series gives
\begin{eqnarray}
z_1 & = & z_0 + R_{56}\delta_0 + T_{566}\delta_0^2 + U_{5666}\delta_0^3 + \ldots \\
\delta_1 & = & \delta_0,
\end{eqnarray}
where the coefficients $R_{56}$, $T_{566}$, $U_{5666}$ etc. are all functions
of the chicane parameters $L_1$ and $\theta_0$. Power series provide a convenient way of
systematically representing transfer maps for beamline components, or sections of beamline.
However, the drawback is that in general, transfer maps can only be represented exactly
by series with an infinite number of terms.  In practice, we have to truncate a power series
map at some order, and the consequence is that we can then lose certain desirable properties
of the map: in particular, a truncated map will not usually be \emph{symplectic}.

\begin{figure}
\begin{center}
\includegraphics[width=0.41\textwidth]{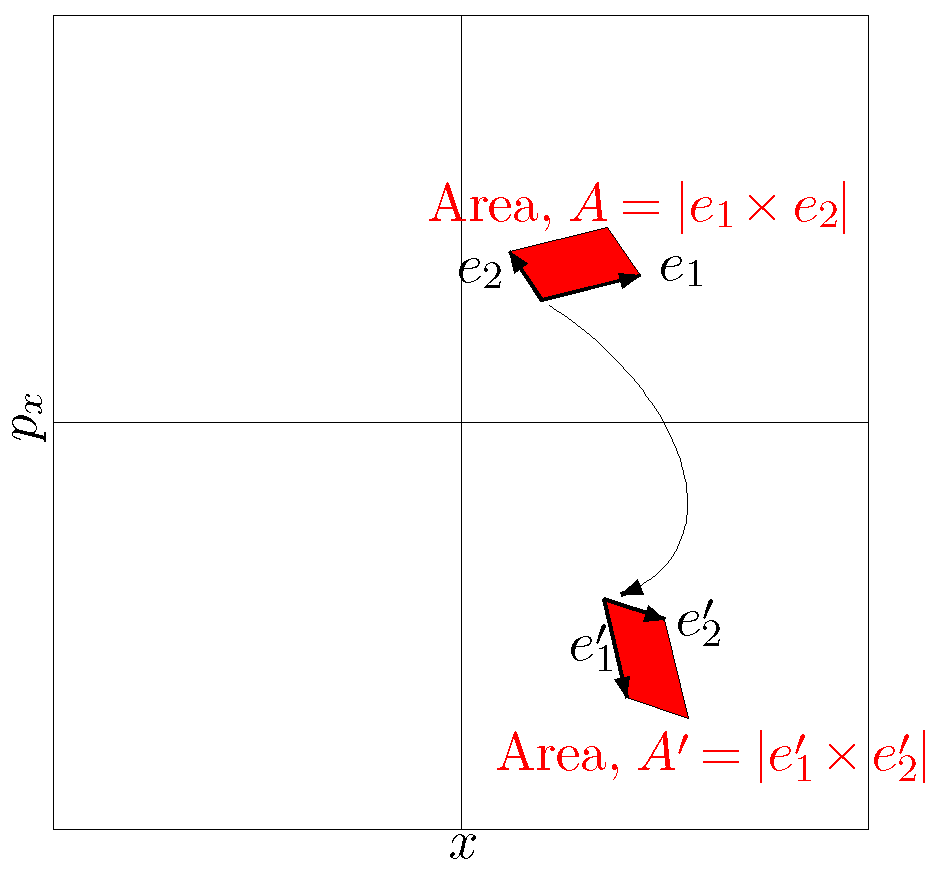}
\end{center}
\caption{Symplectic transformations conserve the area of elements of phase space.
\label{figsymplectictransformation}}
\end{figure}

Mathematically, a transfer map is symplectic if it satisfies the condition:
\begin{equation}
J^\mathrm{T}SJ = S,
\end{equation}
where $J$ is the Jacobian of the map, and $S$ is the antisymmetric matrix with block diagonals:
\begin{equation}
S_2 = \left(
\begin{array}{cc}
0 & 1 \\
-1 & 0
\end{array}
\right).
\end{equation}
The Jacobian of the map is a matrix with elements $J_{mn}$ given by:
\begin{equation}
J_{mn} = \frac{\partial x_{m,f}}{\partial x_{n,i}},
\label{symplecticcondition}
\end{equation}
where $x_{n,i}$ is the $n^\mathrm{th}$ component of the initial phase space
vector $(x, p_x, y, p_y, z, \delta)$ and $x_{m,f}$ is the $m^\mathrm{th}$
component of the final phase space vector.  For a linear map, the elements
of the Jacobian will be constants (i.e.~independent of the phase space variables).
For nonlinear maps, the elements of the Jacobian will depend on the phase space
variables, but if the map is symplectic, then it will still satisfy the condition
(\ref{symplecticcondition}).  Physically, a symplectic transfer map conserves phase space volumes when
the map is applied.  This is Liouville's theorem, illustrated in Fig.~\ref{figsymplectictransformation},
and is a property of charged particles moving in electromagnetic fields, in the absence of radiation
and certain collective effects.

\begin{figure}
\begin{center}
\includegraphics[width=0.35\textwidth]{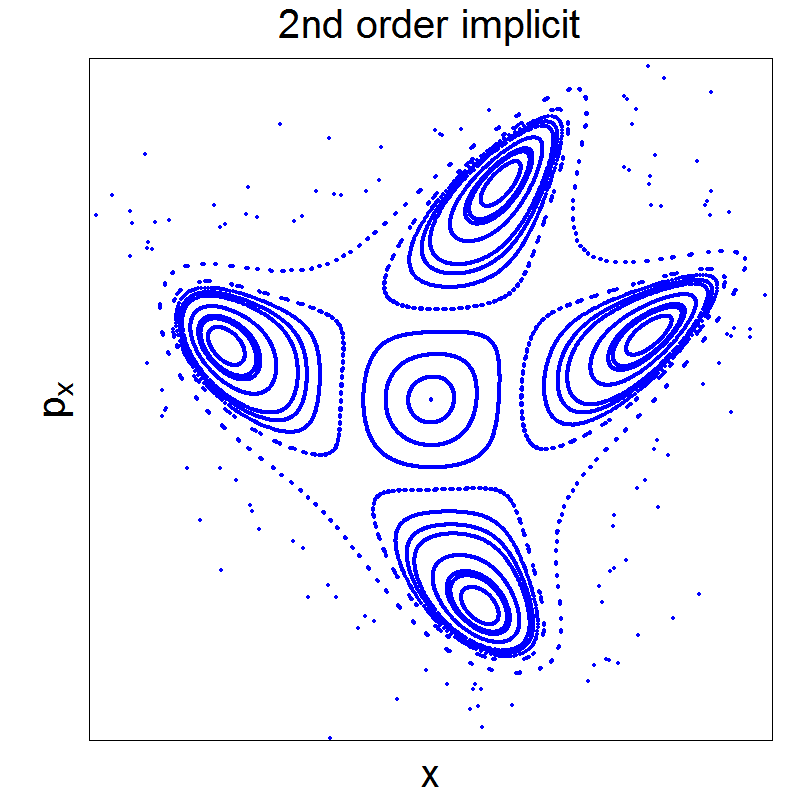}
\includegraphics[width=0.35\textwidth]{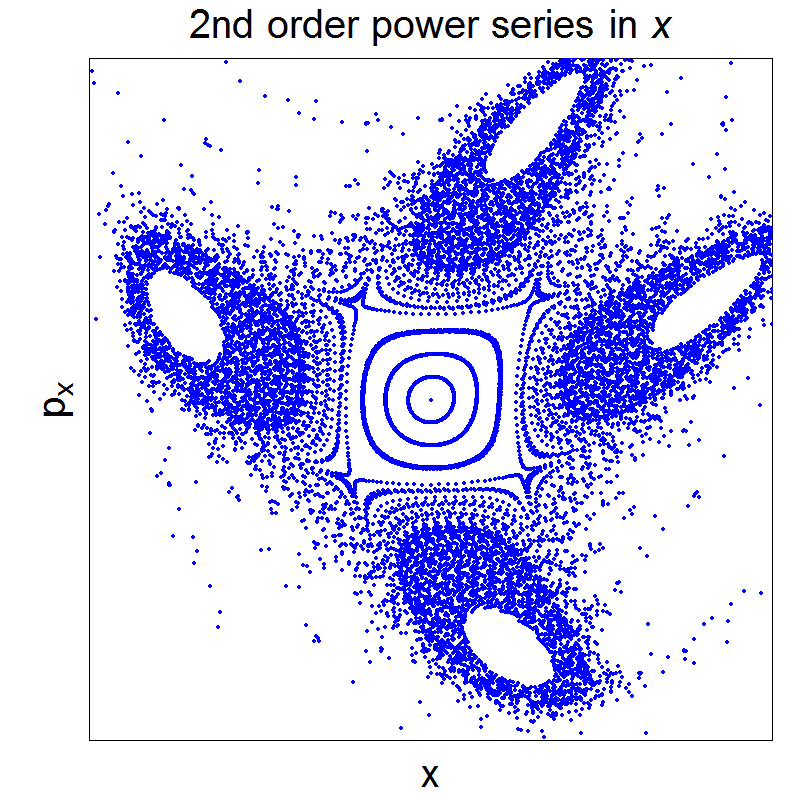}
\end{center}
\caption{Phase space portraits for a simple storage ring containing a sextupole
magnet.  The lattice is the same in both cases, but in the left-hand plot the phase space portrait is constructed using symplectic maps, while non-symplectic maps are used to construct the plot on the right.
\label{symplecticvsnonsymplectic}}
\end{figure}

The effect of losing symplecticity (for example, in the truncation of a power-series map
to finite order) becomes apparent if we compare phase space portraits constructed using
symplectic and non-symplectic transfer maps: an example is shown in
Fig.~\ref{symplecticvsnonsymplectic}.  There are a number of clear differences between
the phase space portraits constructed using symplectic and non-symplectic maps.  Significantly,
the closed loops visible in the ``islands'' in the symplectic case appear blurred when the
non-symplectic map is used.  This is an indication that conserved quantities (constraining
particle motion in the physical system) are maintained by the symplectic map, but not by
the non-symplectic map.  This can have some important consequences for the conclusions
drawn from an analysis of nonlinear effects.  For example, particle trajectories that would be
stable in a real storage ring may appear to be unstable if the storage ring is modelled using
non-symplectic maps, and this can lead to an inaccurate estimate of the dynamic aperture
and the beam lifetime.

\subsection{Symplectic transfer maps}

Real accelerators are often sufficiently complex that it is necessary to construct concise
approximate representations of the transfer maps for the various components: fully
detailed and accurate transfer maps are usually too demanding in terms of computer
processing speed and memory capacity to be of practical use in computational models.
Expanding the transfer map as a power series in the dynamical variables provides a
convenient representation of a transfer map, in terms of a set of coefficients for the power
series.  However, we have seen that the power series that result from an expansion of a
nonlinear transfer map often have an infinite number of terms.  It may be possible in some
cases simply to truncate the power series at some point; however, we have seen that when
we do so, the resulting power series is often not symplectic.  Symplecticity is an important
property of some accelerator systems, and the loss of symplecticity can lead to a model
producing misleading results. 

To address this issue, a number of techniques have been developed for representing
symplectic maps in convenient, concise forms.  For example, it is possible to construct
a power series map that is of finite order, symplectic, and approximates (to a specified
degree) a given symplectic power series map with an infinite number of terms.  This
approach has the additional advantage that a power series representation is \emph{explicit}:
all that is required to apply the map is to substitute particular values of the dynamical 
variables into the given expressions.  However, finite-order symplectic power series are
not easy to construct, and may have limited accuracy unless extended to very high order,
and alternative techniques may be worth investigating.  Examples include the use of
mixed-variable generating functions (see Appendix B) and Lie transformations (Appendix C):
both of these are powerful techniques, but provide \emph{implicit} representations of
transfer maps in the sense that an additional set of equations needs to be solved
(usually numerically) each time the map is applied.  This can make these techniques
cumbersome to implement, computationally expensive, and may limit accuracy.

As an example of a technique for constructing a symplectic map in the form of a
power series, we shall discuss the ``kick'' approximation that is widely used for
modelling multipole magnets in accelerators.  The technique can be applied to
multipoles of any order; but for simplicity, we shall consider just a sextupole magnet.
As usual, and again for simplicity, we consider only motion in one degree of freedom,
though the technique that we develop can readily be extended to include additional
degrees of freedom.

\begin{figure}
\begin{center}
\includegraphics[width=0.7\textwidth]{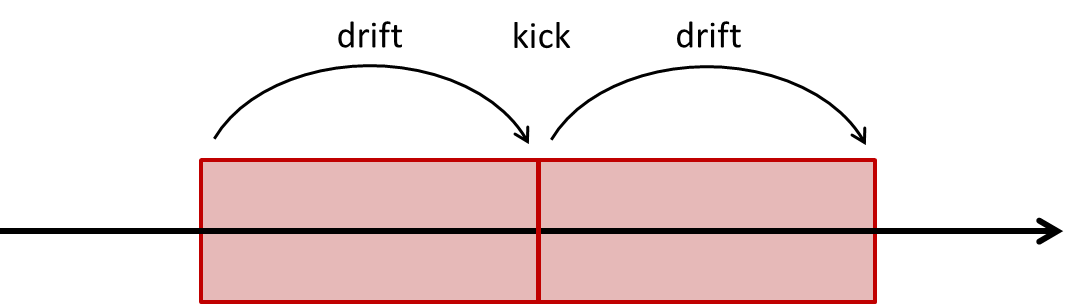}
\end{center}
\caption{Drift--kick--drift approximation for constructing a symplectic transfer map
for a multipole magnet.
\label{driftkickdrift}}
\end{figure}

We start with the equations of motion for a particle moving through the sextupole:
\begin{eqnarray}
\frac{dx}{ds} & = & p_x, \label{sextupoledxds} \\
\frac{dp_x}{ds} & = & -\frac{1}{2} k_2 x^2. \label{sextupoledpxds}
\end{eqnarray}
Since these equations of motion can be derived using Hamilton's equations, the
solution must be symplectic.  Unfortunately, the equations do not have an exact solution
in terms of elementary functions.
However, in the approximation that $p_x$ is constant, we can solve Eq.~(\ref{sextupoledxds});
and in the approximation that $x$ is constant we can solve Eq.~(\ref{sextupoledpxds}).
We therefore split the integration into three steps, making the one or the other approximation at
each step.  If the total length of the magnet is $L$, we take the first step over $0 \le s < L/2$,
making the approximation that $p_x$ is constant, so that the transfer map is:
\begin{equation}
x_1 = x_0 + \frac{L}{2} p_{x0}, \quad p_{x1} = p_{x0}. \label{driftkickdrift1}
\end{equation}
Then we make the approximation that $x$ is constant, taking $x = x_1$, and integrate
Eq.~(\ref{sextupoledpxds}) over the full length of the magnet:
\begin{equation}
x_2 = x_1, \quad p_{x2} = p_{x1} - \frac{1}{2} k_2Lx_1^2. \label{driftkickdrift2}
\end{equation}
Finally, we again make the approximation that $p_x$ is constant, and integrate
Eq.~(\ref{sextupoledxds}) over $L/2 < s \le L$:
\begin{equation}
x_3 = x_2 + \frac{L}{2} p_{x2}, \quad p_{x3} = p_{x2}. \label{driftkickdrift3}
\end{equation}
Because of the way that we have approximated the equations of motion, each of
the transfer maps expressed in Eqs.~(\ref{driftkickdrift1}), (\ref{driftkickdrift2})
and (\ref{driftkickdrift3}) is symplectic.
The (approximate) solution (\ref{driftkickdrift1})--(\ref{driftkickdrift2}) to the
equations of motion (\ref{sextupoledxds}) and (\ref{sextupoledpxds}) is an example of
a \emph{symplectic integrator}.  For obvious reasons, this particular integrator
is known as a ``drift--kick--drift'' approximation: see Fig.~\ref{driftkickdrift}.
In this case, the approximation can work well if the length of the sextupole is short
(compared to the betatron wavelength).  However, by splitting the integration into
smaller steps, it is possible to obtain better approximations.  Using techniques from
classical mechanics, it can be shown that by splitting a multipole in particular ways,
that is with certain combinations of drifts of varying lengths and kicks of different
strengths, it is possible to minimise the error for a given number of integration steps:
simply dividing a multipole into steps of equal length and applying kicks of equal strength
is not usually the best solution.

\subsection{Analysis techniques for nonlinear dynamics}

Power series maps are useful for particle tracking, but do not give much insight
into the dynamics of a given nonlinear system.  To develop a deeper understanding
(for example, to determine the impact of individual resonances in a storage ring) more
powerful techniques are needed.  There are two approaches that are widely used in
accelerator physics: perturbation theory \cite{goldstein2013,ruth1986a,ruth1986b},
and normal form analysis \cite{dragtfinn1979,neridragt1988,forest1998}.  In both these
techniques, the goal is to construct a quantity that is invariant under application of
the transfer map.  Unfortunately, in both cases the mathematics is complicated and
fairly cumbersome, and we do not discuss the details here.  However, by way of
illustration, consider the case of a simple storage ring containing a
single sextupole in each periodic cell.  Normal form analysis provides an expression
for the betatron action $J_x$ of a particle in terms of the phase $\phi_x$ in this case
\cite{wolski2014a}:
\begin{equation}
J_x \approx I_0 - \frac{k_2L}{8}(2\beta_x I_0)^{3/2}
\frac{\left(\cos(3\mu_x/2+2\phi_x) + \cos(\mu_x/2)\right)}{\sin(3\mu_x/2)} + O(I_0^2),
\label{normalformjx}
\end{equation}
where $I_0$ is a constant (an invariant of the motion), $\phi_x$ is the
angle variable, and $\mu_x$ is the phase advance per cell.
Recall that the cartesian variables $x$, $p_x$ are related to the action--angle
variables $J_x$, $\phi_x$ through Eqs.~(\ref{xfromjphi}) and (\ref{pxfromjphi}).

Note that the second term in the expression (\ref{normalformjx}) for $J_x$ becomes very
large when $\mu_x/2\pi$ is close to a third integer: this is the indication of a resonance.  For
linear motion, the action will be constant.  The sextupoles in the lattice introduce some
nonlinearity, which leads to a variation of the action as a particle moves around the ring
(i.e.~as the phase $\phi_x$ increases).  The variation in the action becomes very large
close to a third-order resonance.  Although higher-order resonances may also be driven
by the sextupoles, these are not shown in the expression in Eq.~(\ref{normalformjx}),
which is based on normal form analysis carried out only up to a given order.

\begin{figure}
\begin{center}
\begin{tabular}{c}
phase advance $\mu_x = 0.28\times 2\pi$ \\
\includegraphics[width=0.78\textwidth]{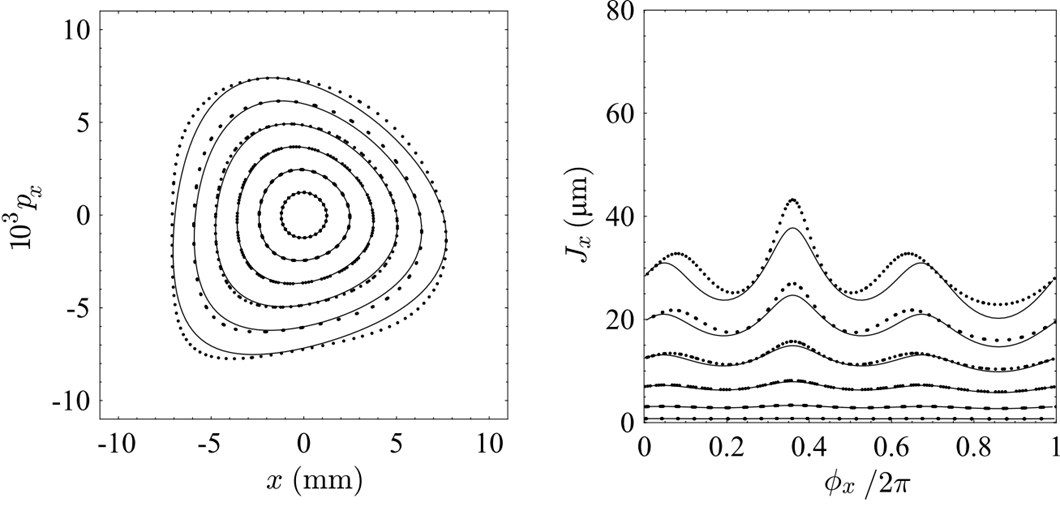}\vspace*{0.1in} \\
phase advance $\mu_x = 0.30\times 2\pi$ \\
\includegraphics[width=0.78\textwidth]{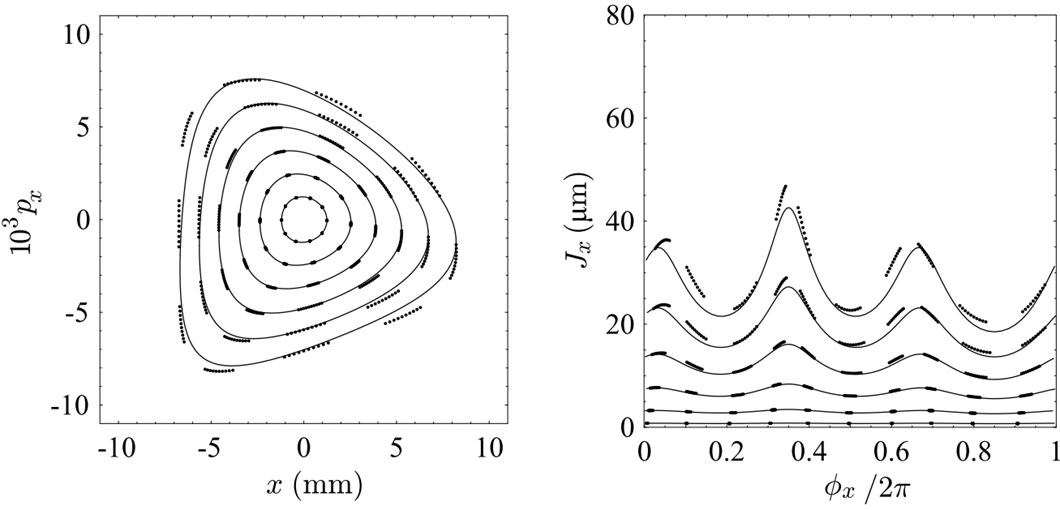}\vspace*{0.1in} \\
phase advance $\mu_x = 0.315\times 2\pi$ \\
\includegraphics[width=0.78\textwidth]{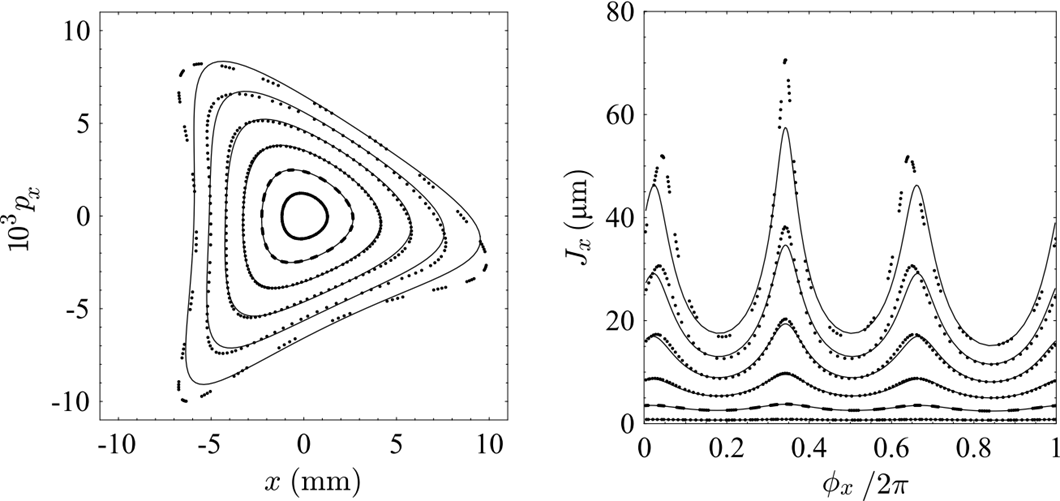}
\end{tabular}
\end{center}
\caption{Comparison between phase space portraits constructed by particle
tracking, and by application of normal-form analysis.  Each plot shows the horizontal
phase space for a particle moving through a lattice with a single sextupole in each
periodic cell, with phase advance per cell equal to $\mu_x$.  The points are calculated
by particle tracking, using a symplectic integrator.  The lines are constructed from
Eq.~(\ref{normalformjx}), based on normal form analysis.
\label{normalformanalysisillustrated}}
\end{figure}

The phase space distortion (related to the variation in the betatron action as a function
of angle) is illustrated in Fig.~\ref{normalformanalysisillustrated}.  Results from particle
tracking (using a symplectic integrator) are shown as point on the plot; the variation in
the action described by normal form analysis in Eq.~(\ref{normalformjx}) is shown as
a solid line.  Although there is not exact agreement, the normal form analysis does give
a reasonable description of the dynamics, at least at low amplitude.  Very large distortions
(resulting from motion close to a resonance, or from a strongly-driven resonance, or at
large amplitude) are difficult to describe analytically.

\subsection{Tune shift with amplitude}

Close inspection of the plots in Fig.~\ref{normalformanalysisillustrated} reveals another
effect, in addition to the obvious distortion of the phase space ellipses:
the phase advance per turn (i.e.~the tune) varies with increasing
betatron amplitude.  This effect is known simply as \emph{tune shift with amplitude}.
Normal form analysis and perturbation theory can both be used to obtain
estimates for the size of the tune shift with amplitude, in terms of the coefficients
of a series expansion for the tune in terms of the action:
\begin{equation}
\nu_x = \nu_{x0} + \left. \frac{\partial \nu_x}{\partial J_x} \right|_{J_x = 0} J_x + 
\frac{1}{2} \left. \frac{\partial^2 \nu_x}{\partial J_x^2} \right|_{J_x = 0} J_x^2 + \ldots
\label{tuneshiftexpansion}
\end{equation}
Note that in general, the tune in each plane (transverse horizontal and vertical) depends
not just on the action in the corresponding plane, but also on the action in the opposite plane.
One consequence of the fact that the motion (under appropriate conditions) is symplectic is
that the horizontal tune shift with vertical amplitude is equal, or at least very close to,
the vertical tune shift with horizontal amplitude.

\begin{figure}
\begin{center}
\begin{tabular}{cc}
$\mu_x = 0.330\times 2\pi$ & $\mu_x = 0.336\times 2\pi$ \\
\includegraphics[width=0.45\textwidth]{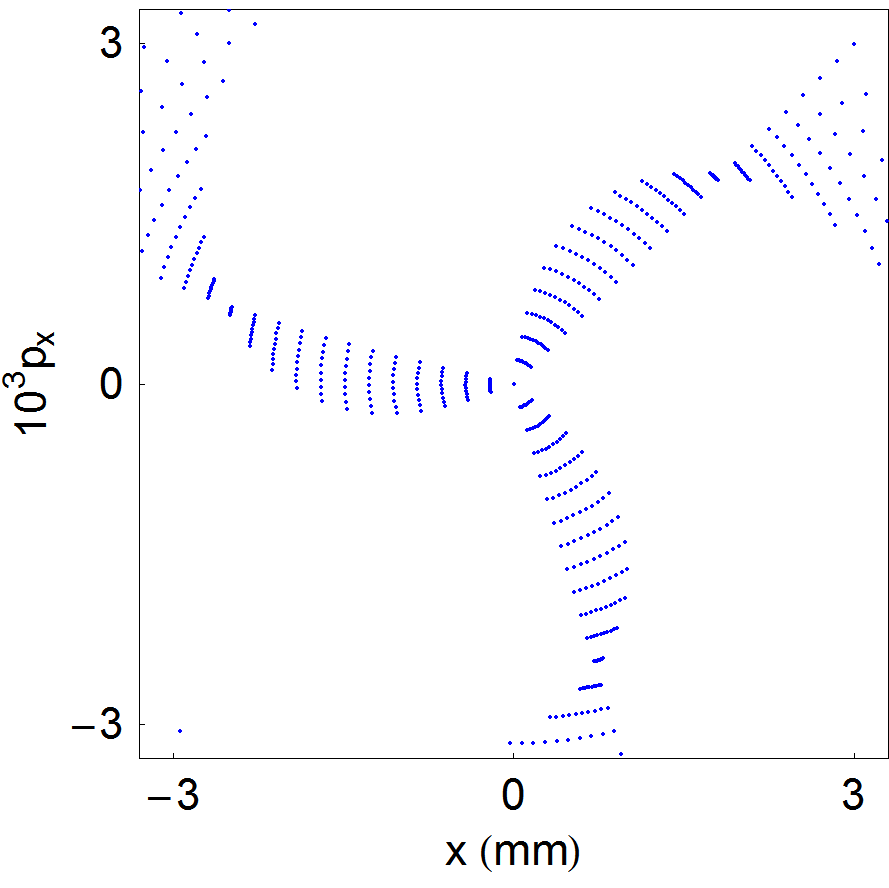} &
\includegraphics[width=0.45\textwidth]{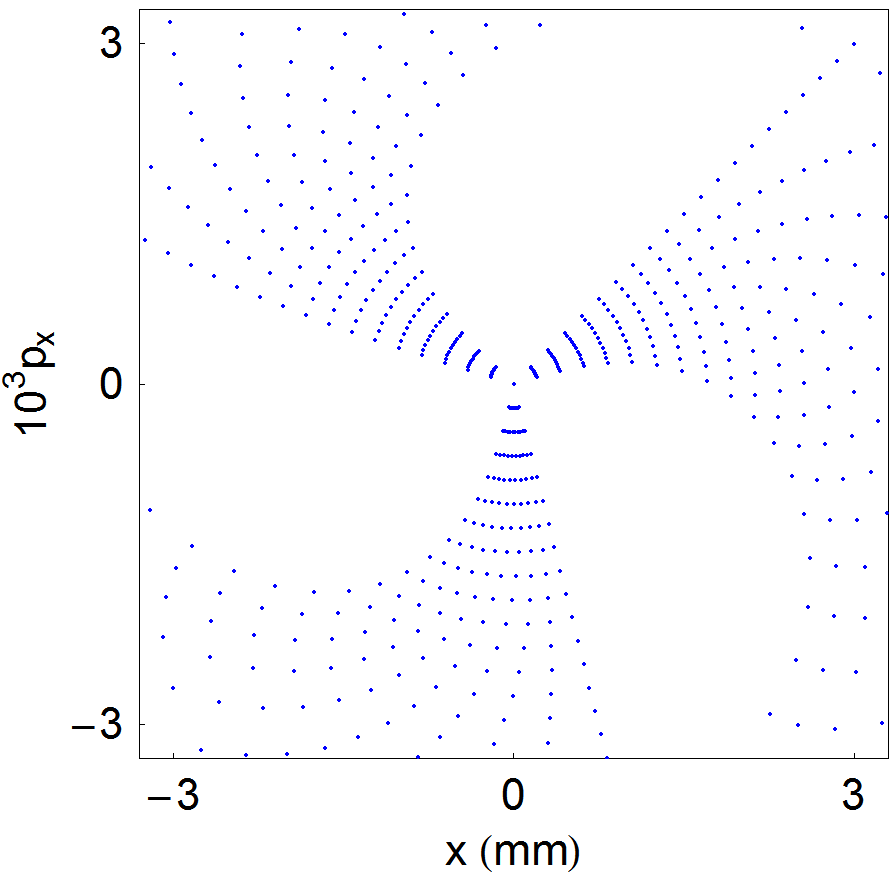}
\end{tabular}
\end{center}
\caption{Tune shift with amplitude: the phase space portraits are constructed by tracking
particles with different betatron amplitudes through 30 periodic cells, with a single nonlinear
element (an octupole) in each cell.  The phase advance per cell varies depending on the
betatron amplitude, and on the phase advance per cell in the limit of zero amplitude (shown
above each plot).
\label{tuneshiftphasespaceportrait}}
\end{figure}

Where the nonlinearity in a storage ring arises from a single sextupole in each periodic cell,
the lowest order term in the expansion in Eq.~(\ref{tuneshiftexpansion}) is second-order in the action.
An octupole, however, does have a first-order tune shift with amplitude, given by:
\begin{equation}
\nu_x = \nu_{x0} + \frac{k_3L\beta_x^2}{16\pi}J_x + O(J_x^2),
\end{equation}
where $k_3L$ is the integrated strength of the octupole, integrated over the length of the octupole:
\begin{equation}
k_3L = \frac{q}{P_0} \int_0^{L} \frac{\partial^3 B_y}{\partial x^3} \, ds.
\end{equation}
The tune shift with amplitude in a storage ring with a single octupole per cell becomes obvious
if we construct a phase space portrait by tracking particles through a small number of cells: an
example (for 30 cells) is shown in Fig.~\ref{tuneshiftphasespaceportrait}.

Tune shift with amplitude helps to explain the ``islands'' that we observed appearing in the
phase space portraits for a storage ring with a single sextupole per periodic cell, see
Fig.~\ref{ps1}.  The islands (closed loops around points away from the origin in phase space)
are associated with resonances, and appear at amplitudes where the phase advance is $2\pi$
times a simple ratio of two integers, and where the transfer map contains an appropriate driving
term for a resonance at the corresponding phase advance.  The number of islands is
determined by the denominator of the ratio of integers, and the width of the islands depends
on the size of the tune shift with amplitude, and on the strength of the driving term.

\begin{figure}
\begin{center}
\begin{tabular}{cc}
\includegraphics[trim=0 10 0 10,clip,width=0.49\textwidth]{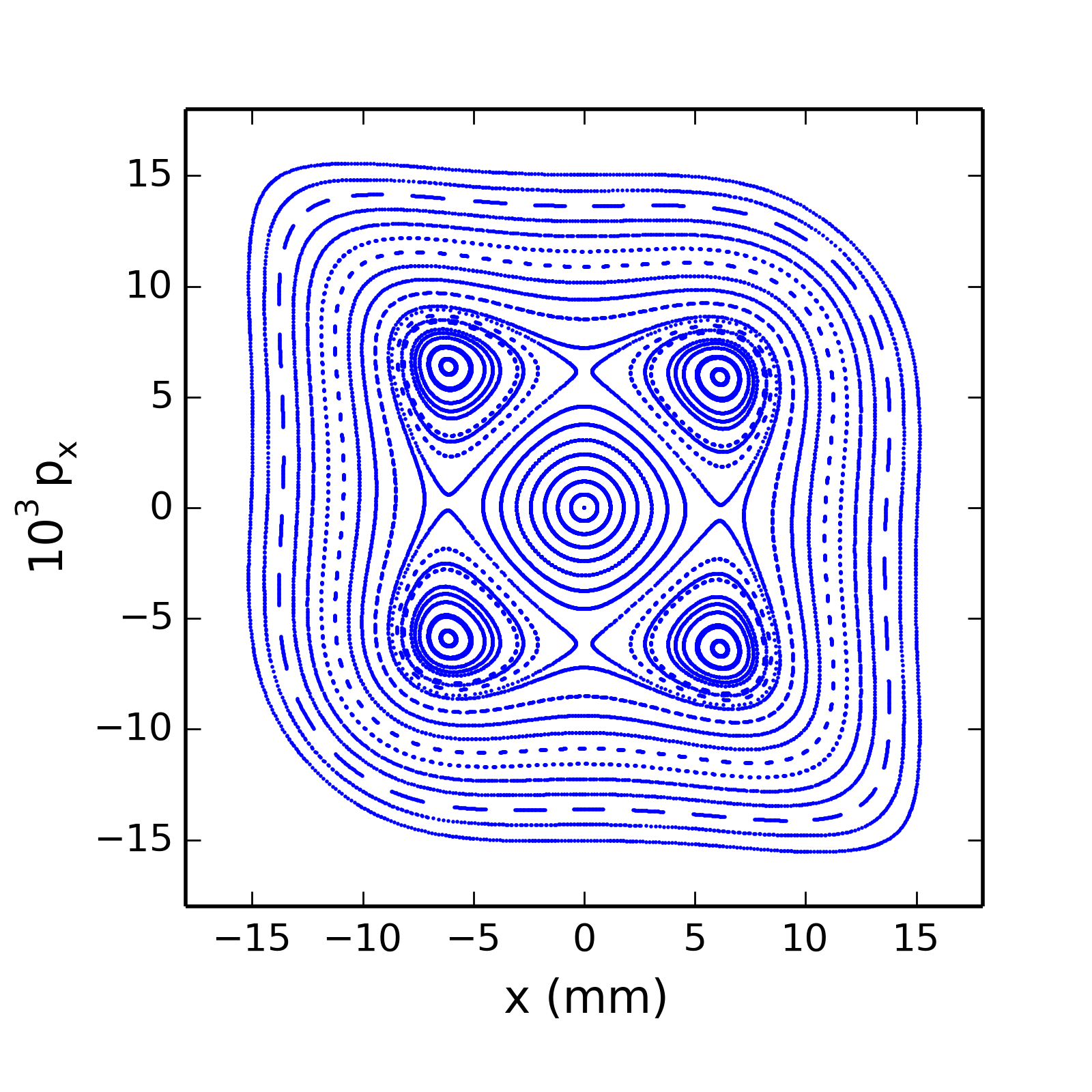} &
\includegraphics[trim=0 10 0 10,clip,width=0.49\textwidth]{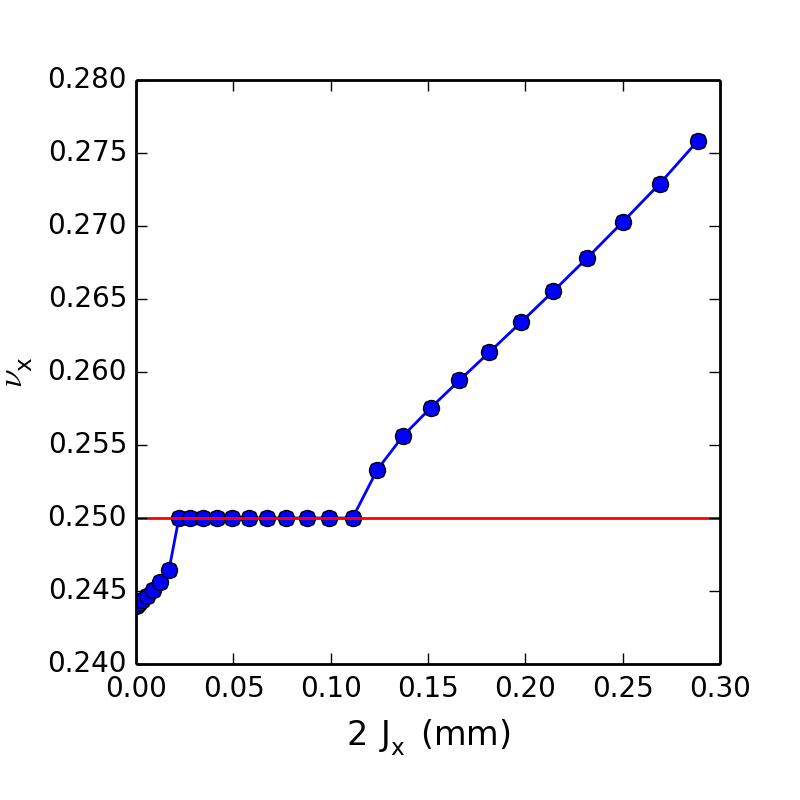}
\end{tabular}
\end{center}
\caption{Phase space portrait (left) and horizontal tunes as a function of the initial particle action (right) for a one-turn map consisting of an octupole magnet and a linear map tuned to a phase advance of $\mu_x=0.244\times 2\pi$.
\label{octupole4thorder}}
\end{figure}

\begin{figure}
\begin{center}
\begin{tabular}{cc}
\includegraphics[trim=0 10 0 10,clip,width=0.49\textwidth]{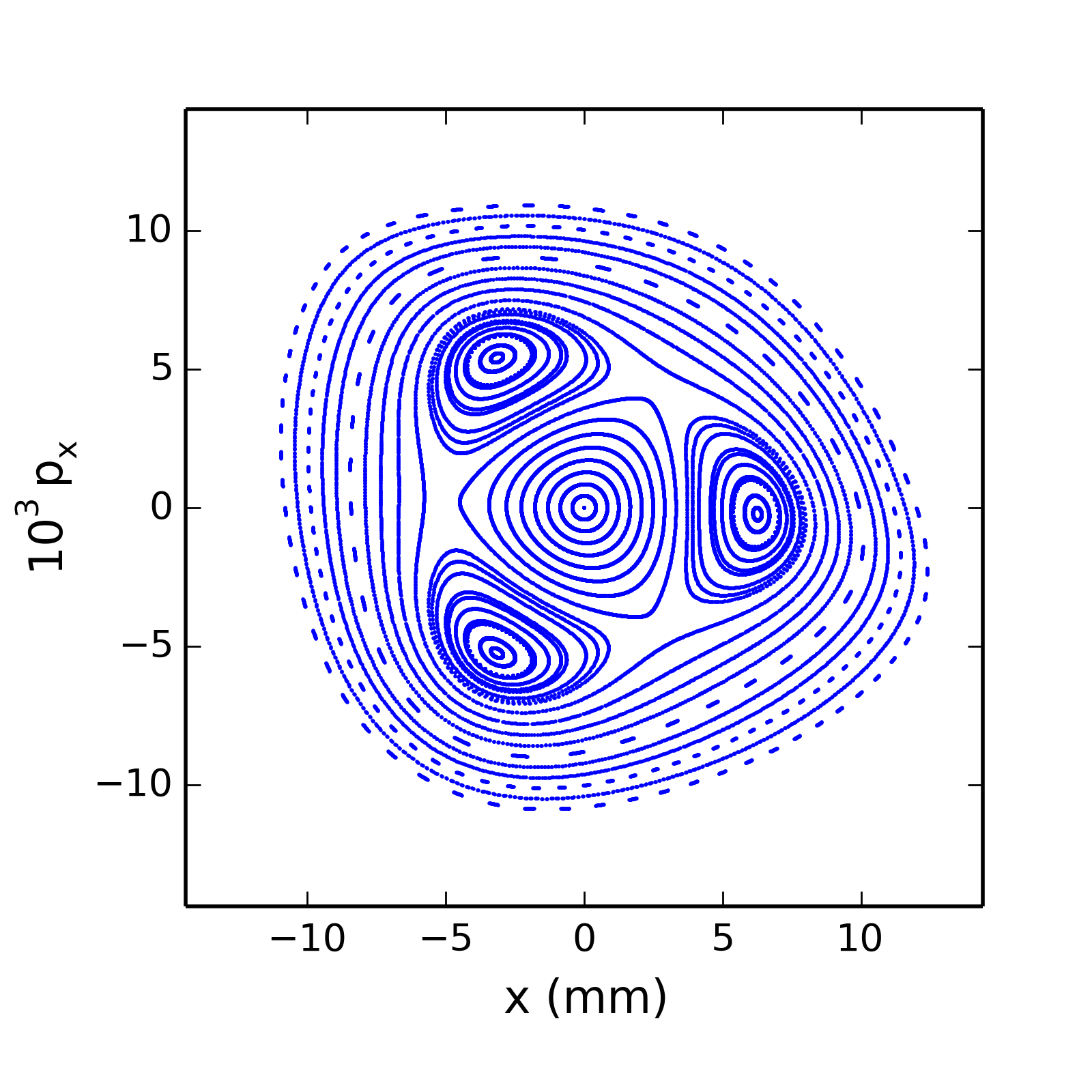} &
\includegraphics[trim=0 10 0 10,clip,width=0.49\textwidth]{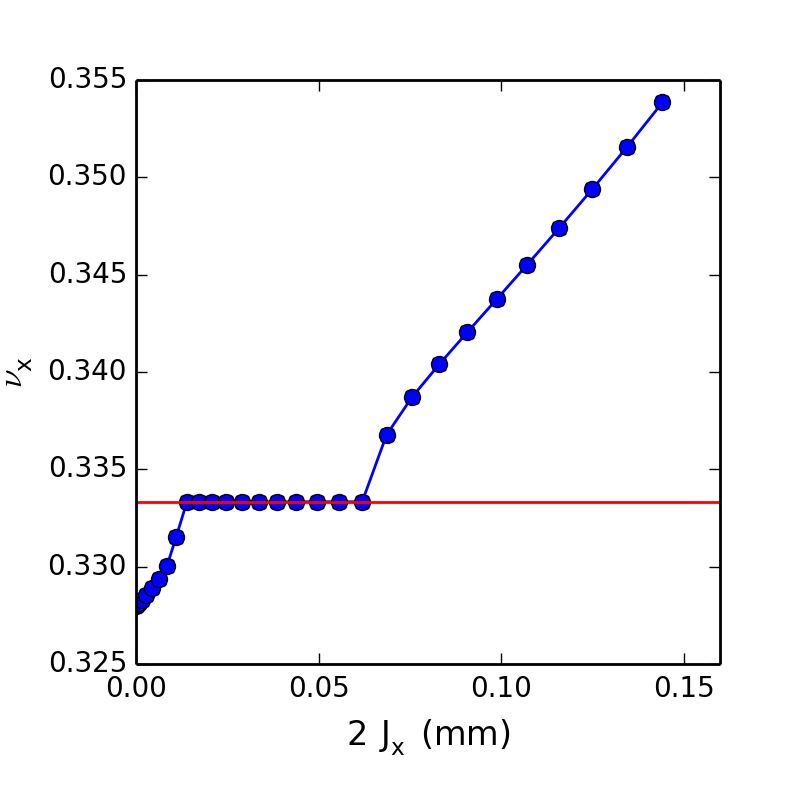}
\end{tabular}
\end{center}
\caption{Phase space portrait (left) and horizontal tunes as a function of the initial particle action (right) for a one-turn map consisting of a sextupole magnet, an octupole magnet and a linear map tuned to a phase advance of $\mu_x=0.328\times 2\pi$.
\label{sextupoleoctupole3rdorder}}
\end{figure}

For example, if we consider a one-turn map with a single octupole magnet and a linear map
tuned for a phase advance close to a $4^\textrm{th}$ order resonance, we observe four stable islands 
in the phase space as shown in Fig.~\ref{octupole4thorder} (left). The tune shift with amplitude is 
illustrated in Fig.~\ref{octupole4thorder} (right), showing the tunes of the tracked particles as a 
function of their initial action ($2J_x$). Particles with small amplitude have tunes below the resonance. 
For large amplitudes, particles have tunes above the resonance and their tunes increase linearly 
with action. Particles trapped in the resonances islands have (on average) a tune locked to the 
resonance condition. As mentioned above, the width of the resonance islands depends on the
tune shift with amplitude and on the resonance strength. This is also demonstrated in another
case, where we consider a one-turn map consisting of a sextupole, an octupole and 
a linear map tuned to a phase advance of $\mu_x=0.328\times 2\pi$ 
(i.e.~close to the $3^\textrm{rd}$ order resonance) as shown in Fig.~\ref{sextupoleoctupole3rdorder}. 
Here the phase space exhibits three stable islands. Note that the octupole induces a 
large tune shift with amplitude, which stabilises the particle motion at the 
resonance as it moves the particle tune out of the resonance for increasing amplitudes.
For comparison, without the octupole and thus with a much smaller tune shift 
with amplitude, particles get lost at the same resonance as we have seen in Fig.~\ref{ps1}.

\section{Dynamic aperture and Frequency Map Analysis}

The techniques we have mentioned for the analysis of nonlinear effects, perturbation theory
and normal form analysis, are based on the existence of constants of motion in the presence
of nonlinear perturbations.  For linear motion in accelerators, there are constants of motion
given by the action variables; for example, in the horizontal plane:
\begin{equation}
2J_x = \gamma_x x^2 + 2\alpha_x x p_x + \beta_x p_x^2.
\end{equation}
In a periodic lattice, the Courant--Snyder parameters have the same periodicity as the lattice
itself, and the action $J_x$ remains constant as a particle moves along the beamline.  When
nonlinear components are present (e.g.~sextupoles) then the betatron action can vary
with position; but normal form analysis may still identify constants of the motion.  An example
is the quantity $I_0$ in Eq.~(\ref{normalformjx}).

It is not obvious, however, that constants of motion can exist in the presence of nonlinear
perturbations.  The fact that they can is a consequence of the \emph{Kolmogorov--Arnold--Moser
(KAM) theorem} \cite{arnold1989}.  The KAM theorem expresses the general conditions for the existence of
constants of motion in nonlinear Hamiltonian systems, and is of particular significance in
accelerator beam dynamics since it tells us that resonances do not invariably result in the
immediate loss of stability.  Resonances will usually tend to drive the amplitudes of particles
with a particular tune to large amplitudes; however, if the tune shift with amplitude is
sufficiently large, then it is possible for there to be a region of stable motion in phase space
at amplitudes significantly larger than that at which resonance occurs.  In simple terms,
the resonance occurs over a limited range of betatron amplitudes: at lower or larger amplitudes,
the tune shift with amplitude moves the particle motion away from the resonance, meaning
that the motion can again be stable. However, the overlapping of two resonances is associated
with a transition from regular to chaotic motion, which is certainly associated with instability.
The parameter range over which the particle motion becomes chaotic is described by the
\emph{Chirikov criterion} \cite{chirikov1979}.

We have so far focused on motion in one degree of freedom.  In that case (for example, in
the phase space portraits shown in Fig.~\ref{ps1}) instability occurred when the oscillation
amplitude exceeded a certain value.  In multiple degrees of freedom, a new phenomenon
occurs: \emph{Arnold diffusion} \cite{arnold1964}.  This refers to the fact that there can be regions of phase space
where motion with large amplitude is stable (associated with the existence of constants of the
motion), while regions of chaotic motion exist at \emph{smaller} amplitudes.  In storage rings,
for example, this means that particle trajectories with initially small amplitudes can be unstable,
even if trajectories with much larger amplitudes are stable.

\begin{figure}
\begin{center}
 \includegraphics[width=0.69\textwidth]{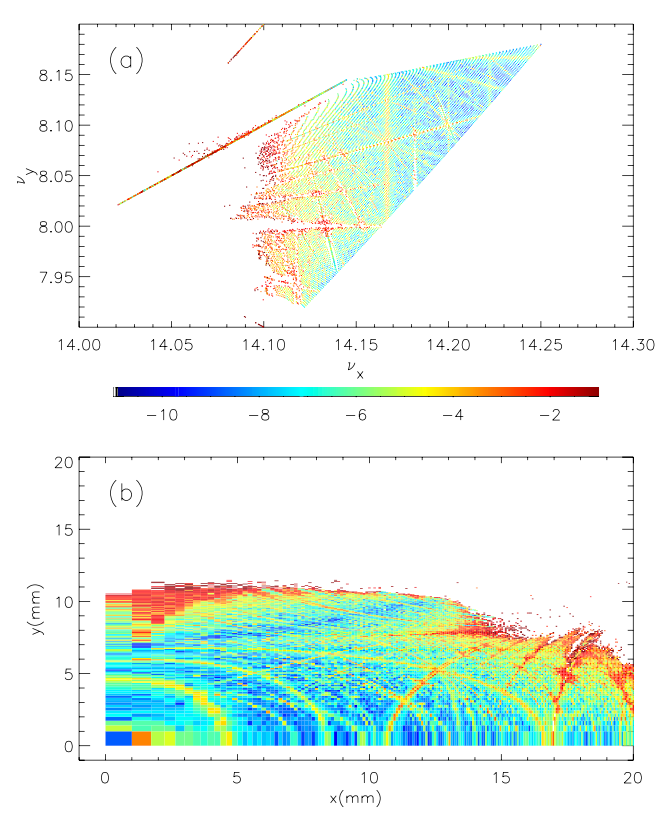}
\end{center}
\caption{Example of frequency map analysis for the Advanced Light Source at Lawrence Berkeley National Laboratory, obtained from analysis of particle tracking data \cite{laskar2003}.
\label{fma_als}}
\end{figure}

One way of studying Arnold diffusion in dynamical systems is through the use of
\emph{frequency map analysis} (FMA) \cite{laskar1990, laskar1994}.  This applies a technique
known as ``numerical analysis
of the Fourier frequencies'' to particle tracking data, either from a simulation or collected experimentally,
to determine with high precision the tunes associated with different particle trajectories.
The strengths of different resonances can be seen by plotting points in
tune space, with diffusion rates shown by different colours: an example for the Advanced Light Source 
(ALS) at Lawrence Berkeley National Laboratory (LBNL) is shown in Fig.~\ref{fma_als}. The FMA
reveals several properties of the nonlinear dynamics in a very visual way, such as 
the detuning with amplitude, the onset of chaos in co-ordinate space and the corresponding
areas in tune space, resonances crossed by the tune footprint, areas of large diffusion, and so on. 
The boundary of the stable region in co-ordinate space is known as the
\emph{dynamic aperture}.

A large dynamic aperture is needed both for good efficiency of injection of the beam
into a storage ring, and for good lifetime of the stored beam.  Although the beam may be
much smaller than the dynamic aperture, scattering processes within the beam can result
in particles acquiring large amplitude betatron or synchrotron oscillation amplitudes, which
may take them outside the dynamic aperture.  When that happens, the particles will be lost
from the machine. Achieving sufficiently large dynamic aperture is typically one of the biggest
challenges in the design of modern light sources, due to their inherent strong nonlinearities.

\section{Summary and further reading\label{summary}}

The effects of nonlinear dynamics impact a wide variety of accelerator systems, including
single-pass systems (such as bunch compressors) and multi-turn systems (such as storage
rings).  It is possible to model nonlinear dynamics in a given component or section of
accelerator beamline by representing the transfer map as a power series.  Power series
provide a convenient (explicit) representation of a transfer map for modelling nonlinear
effects and and for simple analysis of nonlinear dynamics in accelerators.  However,
nonlinear transformations associated with particular accelerator components can usually
only be represented with complete accuracy by power series with an infinite number of
terms: in practice, it is necessary to truncate the power series, i.e.~to drop terms above
some order in the dynamical variables.

Conservation of phase space volumes, expressed in Liouville's theorem, is an important feature
of the beam dynamics in many systems; one example is the conservation of the beam emittances
(in the absence of synchrotron radiation and certain collective effects). To conserve phase
space volumes, transfer maps must be symplectic; but in general, truncated power series maps
are not symplectic.  There are alternative representations that guarantee symplecticity, but
these representations are usually less convenient.  For example, while power series maps are
explicit in that they require only the substitution of values into given expressions, symplectic
maps are often implicit, requiring the numerical solution of a set of equations each time they
are applied.  Techniques do exist, however, that allow the construction of a symplectic transfer map in
an explicit form.  An example is the ``drift--kick--drift'' approximation for a multipole magnet.
A map constructed using such a technique is known as an explicit symplectic integrator.

Common features of nonlinear dynamics in accelerators include phase space
distortion, tune shifts with amplitude, resonances, and instability of particle
trajectories at large amplitudes (limits on dynamic aperture and energy acceptance).
Analytical methods such as perturbation theory and normal form analysis
can be used to estimate the impact of nonlinear perturbations in terms
of quantities such as resonance strengths and tune shift with amplitude.
Analytical studies are often supported by tracking simulations and by numerical
techniques, such as frequency map analysis, that can provide powerful tools for
characterising effects of nonlinear dynamics in accelerators, including tune shifts
and resonance strengths.  Understanding (and correcting, where necessary and possible)
nonlinear effects in accelerators is important for optimising their performance, for example
in achieving a good beam lifetime in a storage ring.

As we mentioned in the introduction, there are many publications that cover the material
discussed in these notes.  Linear optics, and some of the general principles and techniques
of nonlinear dynamics, are covered in (for example) \cite{sylee2011, wolski2014, wiedemann2015}.
Many of the tools of nonlinear dynamics are based on standard techniques in classical
mechanics: this includes the use of mixed-variable generating functions for the representation
of symplectic maps, and canonical perturbation theory.  Such methods are widely covered in
standard text on classical mechanics, for example \cite{goldstein2013, handfinch1999}.
Perturbation theory
applied to accelerator physics is discussed in \cite{ruth1986a, ruth1986b, peggstalman1986}.
Normal form analysis has proved a powerful tool in many situations, and is developed in
\cite{dragtfinn1979, neridragt1988, forest1998}.  Symplectic integration is an important
topic in nonlinear dynamics in accelerators, and is discussed in
\cite{ruth1983, forest2006, forestruth1990, yoshida1990}; a more general text is
\cite{hairer2006}.  Frequency map analysis is reviewed and discussed in
\cite{laskar1994, laskar1995, laskar2003, papaphilippou2014}.

\appendix

\section{Mixed-variable generating functions}

A mixed-variable generating function represents a transfer map (or, more
generally, a canonical transformation) in the form of a function of initial
and final values of the phase space variables.  There are different kinds of
generating function. A mixed-variable generating function of the third kind
(in Goldstein's nomenclature \cite{goldstein2013}) is expressed as a
function of the initial momenta $\vec{p}$ and final co-ordinates $\vec{X}$
(i.e.~the momenta before applying the transformation, and the co-ordinates
after applying the transformation):
\begin{equation}
F_3 = F_3(\vec{X}, \vec{p}).
\end{equation}
The final momenta $\vec{P}$ and initial co-ordinates $\vec{x}$ are obtained by:
\begin{equation}
\vec{x} = -\frac{\partial F_3}{\partial \vec{p}}, \quad \textrm{and} \quad
\vec{P} = -\frac{\partial F_3}{\partial \vec{X}}.
\label{mvgf3equations}
\end{equation}

As an example, consider the mixed-variable generating function in one degree of freedom:
\begin{equation}
F_3 = -X p_x + \frac{1}{2}L p_x^2.
\label{mvgfdrift}
\end{equation}
Applying (\ref{mvgf3equations}) leads to the equations:
\begin{equation}
x = X-L p_x, \quad \textrm{and} \quad
P_x = p_x.
\end{equation}
In this case, the equations are easily (trivially!) solved to give explicit expressions for
$X$ and $P_x$ in terms of $x$ and $p_x$:
\begin{equation}
X = x + L p_x, \quad \textrm{and} \quad
P_x = p_x.
\end{equation}
We see that the function $F_3$ in Eq.~(\ref{mvgfdrift}) is a mixed-variable
generating function for a drift space.
In more general cases, the equations (\ref{mvgf3equations}) need to be
solved numerically each time the transfer map is applied.

\section{Lie transformations}

Lie transformations make use of the fact that the equations of motion for
a particle in an electromagnetic field can be written in the form:
\begin{equation}
\frac{d\vec{x}}{ds} = - :\!H\!: \vec{x}, \label{liealgebraeom}
\end{equation}
where $\vec{x} = (x,p_x, y, p_y, z, \delta)$ is the phase space vector, and
$:\!H\!:$ is a Lie (differential) operator:
\begin{equation}
:\!H\!: = \sum_{i = 1}^3 \frac{\partial H}{\partial x_i}\frac{\partial}{\partial p_i} - \frac{\partial H}{\partial p_i}\frac{\partial}{\partial x_i}.
\label{lieoperator}
\end{equation}
The precise form of the function $H = H(\vec{x})$ (the Hamiltonian) depends
on the field through which the particle is moving.

Formally, a solution to (\ref{liealgebraeom}) can be written:
\begin{equation}
\left. \vec{x} \right|_{s = L} = e^{-L:H:} \left. \vec{x} \right|_{s=0},
\end{equation}
where the exponential of the Lie operator is defined by the usual series expansion
for an exponential:
\begin{equation}
e^{-L:H:} = 1 - L :\!H\!: + \frac{L^2}{2} :\!H\!:^2 - \frac{L^3}{3!} :\!H\!:^3 + \ldots
\end{equation}

The operator $e^{-L:H:}$ is known as a \emph{Lie transformation}.
As a simple example, consider the Hamiltonian:
\begin{equation}
H = \frac{1}{2}p_x^2 + \frac{1}{2}k_1 x^2.
\label{quadrupolehamiltonian}
\end{equation}
Applying the Lie operator $:\!H\!:$ to the variables $x$ and $p_x$ gives:
\begin{eqnarray}
:\!H\!: x & = & -p_x, \\
:\!H\!: p_x & = & k_1 x.
\end{eqnarray}
A further application then gives:
\begin{eqnarray}
:\!H\!:^2 x & = & -k_1 x, \\
:\!H\!:^2 p_x & = & -k_1 p_x,
\end{eqnarray}
and so on.  Then, we find that the Lie transform of $x$ is:
\begin{equation}
e^{-L:H:} x = x + L p_x - \frac{1}{2} k_1 L^2 x - \frac{1}{6} k_1 L^3 p_x + \ldots
\end{equation}
Collecting terms in $x$ and $p_x$, and making use of the series expansions of the
trigonometric functions, we find that:
\begin{equation}
e^{-L:H:} x = x \cos(\sqrt{k_1}L) + p_x \frac{\sin(\sqrt{k_1}L)}{\sqrt{k_1}}.
\end{equation}
Similarly, we find:
\begin{equation}
e^{-L:H:} p_x = p_x - k_1 L x - \frac{1}{2} k_1 L^2 p_x + \frac{1}{6} k_1^2 L^3 x + \ldots
\end{equation}
which can be written:
\begin{equation}
e^{-L:H:} p_x = p_x \cos(\sqrt{k_1}L) - x \sqrt{k_1} \sin(\sqrt{k_1}L).
\end{equation}
The Hamiltonian in Eq.~(\ref{quadrupolehamiltonian}) leads to the Lie transformation
for a quadrupole (in one degree of freedom).

In the above example, although the Lie transformation led to transformations for the dynamical
variables in terms of infinite series, it was possible to express the series in finite form using
standard trigonometric functions.  Unfortunately, this is not generally the case, and applying
a Lie transformation to a dynamical variable generally leads to an infinite power series that
cannot be written exactly in finite form.  However, the power of Lie transformations lies in
the fact that there are known mathematical rules for combining and manipulating
Lie transformations, and that for any \emph{generator} $g = g(\vec{x})$ the Lie
transformation $e^{:g:}$ represents a symplectic map.  This makes it possible
to use Lie transformations for the analysis of nonlinear dynamics in a given
system (such as a particle accelerator) using finite expressions for the generators,
rather than infinite series.

\end{document}